\documentclass[preprint,12pt]{elsarticle}

\usepackage{amssymb}
\usepackage{amsmath}
\usepackage{algorithm}
\usepackage{algpseudocode}
\usepackage{hyperref}
\usepackage[framemethod=tikz]{mdframed}

\begin{document}

\begin{frontmatter}



\title{Stabilizing the Maximal Entropy Moment Method for Rarefied Gas Dynamics at Single-Precision \tnoteref{tnote1}}%



\tnotetext[tnote1]{The code has been released at \url{https://github.com/scraed/TheMomentGauge}}

\author[1,2]{Candi Zheng}\corref{cor1}
\ead{czhengac@connect.ust.hk}
\cortext[cor1]{Corresponding author}
\author[1]{Yang Wang}
\author[2]{Shiyi Chen}

\address[1]{Department of Mathematics,
  Hong Kong University of Science and Technology, Clear Water Bay, Hong Kong SAR, China}
\address[2]{Department of Mechanics and Aerospace Engineering, Southern University of Science and Technology, Xueyuan Rd 1088, Shenzhen, China}

\begin{abstract}
The maximal entropy moment method (MEM) is systematic solution of the challenging problem: generating extended hydrodynamic equations valid for both dense and rarefied gases. However, simulating MEM suffers from a computational expensive and ill-conditioned maximal entropy problem. It causes numerical overflow and breakdown when the numerical precision is insufficient, especially for flows like high-speed shock waves. It also prevents modern GPUs from accelerating MEM with their enormous single floating-point precision computation power. This paper aims to stabilize MEM, making it possible to simulating very strong normal shock waves on modern GPUs at single precision. We improve the condition number of the maximal entropy problem by proposing gauge transformations, which moves not only flow fields but also hydrodynamic equations into a more optimal coordinate system. We addressed numerical overflow and breakdown in the maximal entropy problem by employing the canonical form of distribution and a modified Newton optimization method. Moreover, we discovered a counter-intuitive phenomenon that over-refined spatial mesh beyond mean free path degrades the stability of MEM. With these techniques, we accomplished single-precision GPU simulations of high speed shock wave up to Mach 10 utilizing 35 moments MEM, while previous methods only achieved Mach 4 on double-precision.  
\end{abstract}



\begin{keyword}
Rarefied gas dynamics; Maximal entropy; Machine Learning; Moment Equation; Exponential Family Model; Shock Waves
\end{keyword}

\end{frontmatter}



\section{Introduction}
The transition between dense and rarefied gas dynamics \cite{Kogan1969RarefiedGD} is a typical multiscale problem which is important for understanding gas behavior in extreme conditions \cite{tsien1946superaerodynamics, shizgal1994rarefied} such as in spacecraft re-entry. Such transition is difficult to simulate due to the interaction between two scales: the desired flow motion at macroscopic scale and the non-negligible molecule collisions at mesoscopic scales. Mesoscopic scale simulations, such as the Direct Simulation Monte Carlo (DSMC) \cite{Bird1994} and Discrete Velocity Methods (DVM) \cite{broadwell_1964, DVMbook, mieussens2000discrete, aristov2001direct}, capture molecule collisions hence are accurate for both dense and rarefied gas. However, they are too computationally expensive for dense gas because dense gas has too many colliding molecules. Contrarily, macroscopic scale simulations with coarse-grained resolution in molecule collisions, such as the Navier-Stokes and the Burnett-type equation \cite{agarwal2001beyond, GarcaColn2008BeyondTN}, are computationally efficient for dense gases. They summarize collective behavior of molecules into a few macroscopic quantities such as temperature and use much coarser resolutions in space and time than mesoscopic scale simulations. Nevertheless, such summarization only works for large number of molecules hence is inaccurate for rarefied gas. To efficiently simulate both dense and rarefied gases, a sophisticated balance between resolving molecule collisions and computation costs is required, which involves coarse-graining the key factors for resolving molecule collisions: resolutions in time, space, and molecule velocities.

Works coarse-graining the space and time resolution include the Hilbert-Chapman-Enskog theory \cite{Hilbert1912, Chapman1916,enskog1917kinetische, agarwal2001beyond, GarcaColn2008BeyondTN} and asymptotic preserving methods \cite{mieussens2000discrete, pieraccini2007implicit, dimarco2013asymptotic, bennoune2008uniformly, filbet2010class, xu2010unified, xu2014direct, guo2013discrete, guo2015discrete}. The Hilbert-Chapman-Enskog theory enhances the time resolution of the macroscopic Navier-Stokes equation by perturbative series expansion. However, these methods can exhibit instability \cite{bobylev1982chapman} and may fail to converge \cite{McLennan1965ConvergenceOT}. The asymptotic preserving methods, like UGKS \cite{xu2010unified, xu2014direct} and DUGKS \cite{guo2013discrete, guo2015discrete}, coarse the space and time resolution of the DVM methods utilizing analytical solutions of Boltzmann equations with simplified collision models. Such analytical solutions allow simulations to capture molecule collisions despite the numerical space and time resolution \cite{wang2018comparative}. Recent progress in asymptotic preserving methods has resulted in significantly enhanced efficiency compared to the original DVM or DSMC methods \cite{wang2018comparative}, but they do not address the coarse-graining of molecule velocities.

Works on coarsening the molecule velocities can potentially increase the efficiency further. Such work, including various model equations and moment methods\cite{Grad1949, Struchtrup2011}, aims to approximate molecule's velocities probability distribution by simplified distribution models. They date back to the Mott-Smith model \cite{MottSmith1951TheSO} specifically designed for shock waves and the more general Grad's moment equations \cite{Grad1949} using the Hermite expansion of the Maxwellian distributions. The Grad's moment equations suffer from the loss of hyperbolicity \cite{muller2013rational, Torrilhon2009HyperbolicME} which compromises the well-posedness and stability of hydrodynamic equations. This issue is solved later by the maximal entropy moment method (MEM) \cite{levermore1996moment}, which replaces the Hermite expanded Maxwellian with distributions in the exponential family \cite{dasgupta2011exponential}. Distribution in the exponential family possesses maximized entropy given a particular set of moments and guarantees the hyperbolicity of hydrodynamic equations. Nevertheless, computing such distributions involves the ill-conditioned maximal entropy optimization problem.

The ill-conditioned maximal entropy optimization problem causes numerical overflow and breakdown of the optimization when the numerical precision is insufficient\cite{mcdonald2013affordable, Alldredge2012HighOrderEC, Alldredge2013AdaptiveCO, Schaerer2017The3S}. Previous works tackling the ill-conditioned optimization problem using closed-form approximations \cite{mcdonald2013affordable, giroux2021approximation}, adaptive basis \cite{Abramov2007AnIA, Alldredge2012HighOrderEC, Alldredge2013AdaptiveCO, Schaerer2017The3S}, or regularizations \cite{Alldredge2013AdaptiveCO, alldredge2019regularized}. However, ill-conditionedness still leads to breakdowns for strong non-equilibrium flows \cite{Schaerer2017The3S, Schaerer2017EfficientAA}. It also requires high numerical precision hence preventing modern graphics processing units (GPU) from accelerating the simulation with their enormous (30-80 TFLOPS) single-precision computation power.

This paper tackles the ill-conditionedness of the MEM to prevent possible breakdown and enable single-precision calculation on GPU accelerators. To make the maximal entropy problem less ill-conditioned, we propose gauge transformations that use adaptive Hermite polynomials \cite{Schaerer2017The3S}. It differs from the previous works \cite{Abramov2007AnIA, Alldredge2012HighOrderEC, Alldredge2013AdaptiveCO, Schaerer2017The3S} by transforming the flow field, hydrodynamic equation and numerical solver together into a better coordinate system before formulating the maximal entropy problem, which significantly stabilizes the simulation and accelerates the optimization process. To prevent numerical overflow of the distribution function that causes breakdown, we adopt the canonical form of the exponential family model, which introduces the partition function that can be safely computed using the log-sum-exp trick \cite{blanchard2021accurately}. To enhance the robustness when optimizing entropy with ill-conditioned Hessian, we use Newton's method with a classical Hessian modification that adds a multiple of the identity \cite[p.51]{Nocedal2006}. This modification prevents possible breakdown even for extremely ill-conditioned cases and is simple to implement. Our method succeeded in accurately simulating a Mach 10 normal shock wave using the 35 moments MEM \cite{Schaerer2017The3S} at single precision on a GPU accelerator, while previous work  only reported results at Mach 4 in double precision \cite{Schaerer2017The3S, Schaerer2017EfficientAA}. In addition, we have also demonstrated that MEM has a lower resolution limit on its spatial resolution: the mean free path. Over-refining the spatial mesh beyond this limit is unnecessary and undermines stability. In summary, this paper makes the maximal entropy moment method practical for simulating very strong normal shock waves on modern GPUs at single floating-point precision, with significant stability improvement compared to previous methods.
 
\section{The Maximal Entropy
Moment Method}

We first introduce the Boltzmann equation as the fundamental governing equation for monatomic gas dynamics. Then we show that the maximal entropy moment method (MEM) is an approximation of the Boltzmann equation. Finally, analysis of the approximation error reveal the limitation in spatial resolutions affecting the numerical stability of MEM.

\subsection{The Boltzmann Equation as the Fundamental Governing Equation}\label{subsubsec2}

The fundamental governing equation for monatomic gas dynamics under no external forces is the Boltzmann equation \cite{TheBoltzmannEquationAndItsApplications}\begin{equation} \label{The Boltzmann Equation}
\frac{\partial f}{\partial t} + u_\alpha \frac{\partial f}{\partial x_\alpha} = Q(f,f);\quad \alpha = 1,2,3,
\end{equation}in which $f(t,\bold{x},\bold{u})$ is the one-particle distribution function describing the probability of finding one gas molecule at position $\bold{x}$ with velocity $\bold{u}$ at time $t$, with $u_\alpha$, $x_\alpha$ be the $\alpha$th component of molecule velocity $\bold{u}$ and $\bold{x}$ with the Einstein's summation convention adopted. The term $Q(f,f)$ in the Boltzmann equation is the collision term describing the collision between gas molecules, with $Q$ being a bilinear functional \cite[p.67]{TheBoltzmannEquationAndItsApplications} of the distribution $f$. This paper adopt BGK type collision models for $Q(f,f)$ for efficient calculation, see \ref{sec AB}. The Boltzmann equation characterizes the transportation and binary collision of gas molecules at the mesoscopic scale, which is enough to describe both dense and rarefied gas accurately.

The macroscopic properties of gas, such as number density $n$, flow velocity $\bold{v}$, and temperature $T$, are moments of the one-particle distribution function $f(t,\bold{x},\bold{u})$ w.r.t the molecule velocity $\bold{u}$
\begin{equation}
\begin{split}
        n(t,\bold{x}) &= \int f(t,\bold{x},\bold{u}) d^3 \bold{u}\\ 
    \bold{v}(t,\bold{x}) &= \frac{1}{n(t,\bold{x})}\int \bold{u} f(t,\bold{x},\bold{u}) d^3 \bold{u}\\
      T(t,\bold{x})  &= \frac{m}{3 k_Bn(t,\bold{x})}\int \bold{c}^2 f(t,\bold{x},\bold{u}) d^3 \bold{u},
\end{split}
\end{equation}
in which $m$ is the mass of gas molecules, $\bold{c} = \bold{u} - \bold{v}$ is the peculiar velocity of gas molecules. The mass density $\rho$ used more commonly in hydrodynamics is the product of number density $n$ and molecule mass $m$. 

The one-particle distribution $f$ governed by the Boltzmann equation \eqref{The Boltzmann Equation} is accurate for both dense and rarefied gas but introduces redundant dimensions for molecule velocity $\mathbf{u}$. It is crucial for accurate calculation but unnecessary for observation since we are only interested in the density, flow velocity, and temperature, which are moments of molecule velocities. For efficient computation, it would be beneficial to summarize the one-particle distribution in $\mathbf{u}$ with a few parameters using distribution model functions. 

For example, dense gas at thermodynamic equilibrium could be completely described by its density, flow velocity, and temperature with its one-particle distribution function $f$ in the form of the Maxwell distribution \cite[p.76]{StatisticalPhysicsOfParticles} \begin{equation} \label{The Maxwell distribution}
f_0(t,\bold{x},\bold{u}) = \frac{n}{(2\pi k_b T/m )^{3/2}} \exp(-\frac{m(\bold{u} - \bold{v})^2}{2k_B T} ),
\end{equation}
in which $k_B$ is the Boltzmann constant and we have omitted the dependency of $n, \bold{v}, T$ on time $t$ and spatial coordinate $\bold{x}$ for simplicity. The Maxwell distribution $f_0$ corresponds to the equilibrium gas state at which collisions between molecules are sufficient to make the collision term $Q(f_0,f_0)$ vanish. However, it can not describe non-equilibrium phenomena like viscosity and heat transfer. More accurate computation of gas dynamics requires distribution model functions beyond the Maxwell distribution.


\subsection{The Exponential Family Models for the Distribution Function}\label{subsubsec2}

The maximal entropy moment method aims to model the one-particle distribution function with the exponential family \cite{MAL-001ExpFamily} distribution model. The exponential family distribution has the property that a few moments w.r.t the molecule velocity $\bold{u}$ can completely describe the distribution, similar to the fact that density, flow velocity, and temperature fully determine the Maxwell distribution. This property allows us to summarize the distribution in a few moments without loss of information, thereby eliminating the need to record  distribution values $f(\bold{u})$ with redundant dimensions in molecular velocity $\bold{u}$.

Specifically, the exponential family distribution model record the $f(\bold{u})$ with a few parameters $\alpha_i$ in the form \cite{Torrilhon2016ModelingNG,Schaerer2017The3S, Schaerer2017EfficientAA}
\begin{equation} \label{Maximal Likelihood equation for Exponential Family Distributions: exp family ori}
	\begin{split}
			f(\mathbf{u}; \boldsymbol{\alpha}(t, \bold{x}))&= \exp(\alpha_i(t, \bold{x}) \phi_i(\mathbf{u})); i=0,\cdots, M
	\end{split}
\end{equation}
in which $\{\phi_i(\bold{u}), i=0,\cdots, M\}$ is a set of prescribed sufficient statistics with $\phi_0(\bold{u})=1$, and $\alpha_i$ is the $i$th component of the parameters $\boldsymbol{\alpha}$ with the Einstein's summation convention adopted. These parameters are to be determined according to observations like moments of the distribution. 

Unlike previous works \cite{Torrilhon2016ModelingNG,Schaerer2017The3S, Schaerer2017EfficientAA}, this paper utilizes the canonical form of the exponential family, in which the partition function is crucial in preventing numerical overflow. The canonical form is a different but equivalent notation for the exponential family distribution
\begin{equation} \label{Maximal Likelihood equation for Exponential Family Distributions: exp family}
	\begin{split}
			f(\mathbf{u}; \boldsymbol{\beta}(t, \bold{x}))&= \beta_0(t, \bold{x}) \exp(\beta_i(t, \bold{x}) \phi_i(\mathbf{u})-\log Z(\boldsymbol{\beta};\boldsymbol{\phi})); i=1,\cdots, M\\
			Z(\boldsymbol{\beta};\boldsymbol{\phi})	&=\int\exp(\beta_i(t, \bold{x}) \phi_i(\mathbf{u}))  d^3\mathbf{u}; i=1,\cdots, M,
	\end{split}
\end{equation}
in which $n = \beta_0$ is the number density of gas, $\beta_i$ is the $i$th component of the natural parameters $\boldsymbol{\beta}$ with the Einstein's summation convention adopted, $Z(\boldsymbol{\beta};\boldsymbol{\phi})$ is the partition function. The canonical form and the non-canonical form of the exponential family model are related by
\begin{equation} \label{Maximal Likelihood equation for Exponential Family Distributions: form transformation}
	\begin{split}
			\alpha_0 &= \log( \beta_0 ) - \log(Z(\boldsymbol{\beta};\boldsymbol{\phi})) \\
   \alpha_i &= \beta_i;\quad i\ge 1
	\end{split}
\end{equation}

The choice of sufficient statistics $\phi_i(\bold{u})$ of the exponential family model \eqref{Maximal Likelihood equation for Exponential Family Distributions: exp family} is important to accurately modeling the one-particle distribution function. Such statistics are prescribed according to the prior knowledge of the flow. This paper adopts the 35-moment system \cite{Schaerer2017EfficientAA}, whose sufficient statistics for one-dimensional flow are described in \ref{sec AM35}.

The most important characteristic of the exponential family model \eqref{Maximal Likelihood equation for Exponential Family Distributions: exp family} is that they are the only form of the probability distribution sufficiently described by moments of sufficient statistics
\begin{equation} \label{Maximal Likelihood equation for Exponential Family Distributions: exp family moments}
M_i(t,\bold{x}) = \int  \phi_i(\bold{u})f(\mathbf{u};  \boldsymbol{\beta}(t, \bold{x})) d^3\mathbf{u};i=0,\cdots,M,
\end{equation} according to the Pitman–Koopman–Darmois theorem \cite{Koopman1936OnDA}. It means that moments other than those in \eqref{Maximal Likelihood equation for Exponential Family Distributions: exp family moments} provide no extra information about the exponential family distribution $f(\mathbf{u};  \boldsymbol{\beta}(t, \bold{x}))$. This characteristic justifies summarizing the exponential family distribution into those moments $M_i$ of sufficient statistics.

\subsection{Approximating the One-Particle Distribution Function with Exponential Family Models}\label{subsubsec2}

The one-particle distribution $f$ governed by the Boltzmann equation is not guaranteed to be an exponential family distribution. But we could find its "closest" exponential family approximation $f_{\boldsymbol{\beta}}$ of the form \eqref{Maximal Likelihood equation for Exponential Family Distributions: exp family} via moments of sufficient statistics. The following three equivalent properties lay the theoretical foundation to this.
\begin{itemize}
    \item \textbf{Minimum Distance}: $f_{\boldsymbol{\beta}}$ is the "closest" exponential family distribution to $f$ with the minimized Kullback–Leibler divergence.
    \item \textbf{Moment Matching}: $f_{\boldsymbol{\beta}}$ and $f$ have the same moments of sufficient statistics.
    \item \textbf{Maximal Entropy} $f_{\boldsymbol{\beta}}$ has the maximized entropy among all distributions with moments of sufficient statistics that match $f$.
\end{itemize}
A comprehensive definition of these properties is given in \ref{sec A opt}. These properties indicate that the moments $M_i$ of sufficient statistics uniquely define the parameters $\beta_i$ of $f_{\boldsymbol{\beta}}$.

However, not every configuration of moments \( M_i \) will have associated \( \beta_i \) parameters. Earlier research demonstrated that, for distribution functions in the L1 space on a finite velocity domain, \( \beta_i \) exists if and only if the moments \( M_i \) are realizable \cite{Junk2000MAXIMUMEF, Schaerer2017EfficientAA}. This means the moments \( M_i \) are moments of some L1 distributions functions. Since the moments of gas flow are inherently from the one-particle distribution $f$, it appears that parameters \( \beta_i \) is guaranteed to exist when we confine the distributions to a finite velocity domain.

A caveat of the previous analysis is that it does not considers the particle nature of gas, which is non-negligible for rarefied gas. Gas can be visualized as a collection of discrete points, with a probability density distribution represented by a sum of Dirac-Delta functions. Such distribution does not belong to L1 space, hence their moments may not have corresponding parameters $\beta_i$. 

For instance, in extremely rarefied gases where fewer than $M$ particles exist within a spatial-temporal cell centered at $\mathbf{x}$ and $t$, only $M$ independent moments $M_i(\mathbf{x}, t)$ can be determined. Consequently, with more unknowns ($M+1$ unknown parameters $\beta_i$) than equations, deducing the parameters becomes untenable. A potential solution is to increase the cell size to capture more particles. To ensure the existence of parameters $\beta_i$, there should be a minimum cell size limiting the spatial and temporal resolution of MEM. We'll explore this further in Section \ref{Resolution Limits sec}.

(Response to Reviewer 2 Q2, Q4)

In summary, exponential family distributions allow us to approximate and condense a one-particle distribution function into moments of sufficient statistics. Such an approximation is feasible with appropriate spatial and temporal resolution choices for MEM. However, the one-particle distribution function is not provided until we solve the Boltzmann equation, which is numerically impractical. To obtain the exponential family approximation to the one-particle distribution, we must deduce the maximal entropy moment equations for exponential family distributions as an approximation to the Boltzmann equation.



\subsection{The Governing Equations for Exponential Family Models: Maximal Entropy Moment Equations}\label{Maximal Entropy Moment Equations}

The maximal entropy moment equations are obtained by taking moments of sufficient statistics after substituting $f$ with its exponential family approximation $f_{\boldsymbol{\beta}}$ into the Boltzmann equation. Besides, the collision term $Q$ in the Boltzmann equation could be further simplified using the BGK collision model \eqref{The BGK model}. The resulting maximal entropy moment equations are of the following form:
\begin{subequations}\label{The ME equation}
\begin{align}
    \frac{\partial M_i}{\partial t} +  \frac{\partial F_{i \alpha} }{\partial x_\alpha} &= \frac{\Tilde{M}_i - M_i}{\tau}; i=0,\cdots,M; \alpha=1,2,3 \label{The ME equation:subeq1}\\
    M_{i}(t,\bold{x})&= \int \phi_i(\bold{u}) f_{\boldsymbol{\beta}}(\mathbf{u};\boldsymbol{\beta}(t, \bold{x}))d^3\mathbf{u}\label{The ME equation:subeq2}\\
    F_{i\alpha}(t,\bold{x})&= \int u_\alpha \phi_i(\bold{u}) f_{\boldsymbol{\beta}}(\mathbf{u};\boldsymbol{\beta}(t, \bold{x})) d^3\mathbf{u}\label{The ME equation:subeq3}\\
    \Tilde{M}_{i}(t,\bold{x})&= \int \phi_i(\bold{u}) \Tilde{f}(t, \bold{x}, \mathbf{u}) d^3\mathbf{u}\label{The ME equation:subeq4},
\end{align}
\end{subequations}
in which $\Tilde{f}$ is determined by the BGK model in \ref{sec AB} and we adopt Einstein's summation convention. The maximal entropy moment equations form a closed system of equations because both the moments $M_i$ and the fluxes $F_{i\alpha}$ are functions of the parameters $n$ and $\boldsymbol{\beta}$. Notably, they reduced the redundant dimensions for molecule velocity in the Boltzmann equation by summarizing them into a few moments.


However, the legitimacy of the substitution of \( f \) with \( f_{\boldsymbol{\beta}} \) is questionable. In related literature, this substitution is often viewed as a closure method to truncate the infinite moment equations derived from the Boltzmann equation. Such an interpretation complicates the analysis of MEM's truncation error. Instead, we regard the maximal entropy moment equations as outcomes of the invariant manifold assumption \cite[S2.1]{Gorban2013Hilberts6P}. This perspective clarifies when and why the substitution of \( f \) with \( f_{\boldsymbol{\beta}} \) is valid. Consequently, our approach reveals that MEM serves as a 1st order approximation to the Boltzmann equation. It also uncovers an implicit lower limit on the numerical time step crucial for simulation stability, detailed further in Section \ref{Resolution Limits sec}.

The substitution of \( f \) with \( f_{\boldsymbol{\beta}} \) is valid when the invariant manifold assumption holds. This assumption decomposes the collision term \( Q \) in the Boltzmann equation, capitalizing on its bilinear property, as:
\begin{equation} \label{Collision decompose}
Q(f,f) = Q(f-f_{\boldsymbol{\beta}},f - f_{\boldsymbol{\beta}}) + 2 Q(f-f_{\boldsymbol{\beta}}, f_{\boldsymbol{\beta}}) + Q(f_{\boldsymbol{\beta}},f_{\boldsymbol{\beta}}).
\end{equation}
A similar decomposition could also be achieved for the BGK collision model. Here we only discuss the collision term of the Boltzmann equation without loss of generality. The rightmost term on the right-hand side of \eqref{Collision decompose} steers \( f_{\boldsymbol{\beta}} \) towards equilibrium, nullifying when \( f_{\boldsymbol{\beta}} \) matches a Maxwell distribution. The leading two terms depict the collisional mechanisms drawing \( f \) closer to \( f_{\boldsymbol{\beta}} \), vanishing when \( f \) equals \( f_{\boldsymbol{\beta}} \). Essentially, the leading two terms represent the substitution process.

The invariant manifold assumption states that the leading two terms in the RHS of \eqref{Collision decompose} have very large amplitudes, pulling $f$ to $f_{\boldsymbol{\beta}}$ much faster than the flow's characteristic time scale. In other words, the one-particle distribution $f$ governed by the Boltzmann equation quickly relaxes to its exponential family distribution approximation $f_{\boldsymbol{\beta}}$. The invariant manifold assumption enables us to safely ignore the time required to pull $f$ to $f_{\boldsymbol{\beta}}$, replacing $f$ in the Boltzmann equation directly with $f_{\boldsymbol{\beta}}$ to derive the maximal entropy moment equations.

While the invariant manifold assumption provides a basis for substituting \( f \) with \( f_{\boldsymbol{\beta}} \), its validity is not guaranteed. As a result, the maximal entropy moment equations serve merely as a 1st order approximation to the Boltzmann equation, carrying an inherent approximation error. This intrinsic error is proportional to the time \( \tau_e \) needed for the first two terms to relax \( f \) to \( f_{\boldsymbol{\beta}} \), and this duration can vary based on the chosen sufficient statistics. As discussed in Section \ref{Resolution Limits sec}, the time \( \tau_e \) dictates a minimum threshold for the numerical time resolution in MEM simulations. Beyond this threshold, the intrinsic error of MEM surpasses the numerical error, rendering further increases in resolution superfluous.


\subsection{The Gauge Symmetry of the Exponential Family Model and Gauge transformations}

The exponential family model has inherent freedom in changing the parameter and moments simultaneously, which is its gauge symmetry. It refers to the invariance of the distribution under gauge transformations on sufficient statistics and parameters. Locally, a gauge transformation for exponential family model is an invertible change of sufficient statistics from $\phi_i(\bold{u})$ to another set $\phi'_i(\bold{u})$
\begin{equation}
\label{Gauge basis transformation}
    \phi'(\bold{u})_i = A_{ij} \phi(\bold{u})_j;\quad  i=0,\cdots,M,
\end{equation}
in which $A_{ij}$ is an arbitrary invertible matrix. This gauge transformation on sufficient statistics induces transformation on moments, fluxes and parameters as
\begin{equation}
\label{Gauge transformation 2}
M'_i = A_{ij} M_j\quad \alpha'_i = (A^T)^{-1}_{ij} \alpha_j \quad  F'_{i\alpha} = A_{ij} F'_{j\alpha}\quad i=0,\cdots,M
\end{equation}
in which $F_{i\alpha}$ is the flux of the moment $M_i$ defined in \eqref{The ME equation}, $\alpha_i$ is the parameters of the exponential family model defined in \eqref{Maximal Likelihood equation for Exponential Family Distributions: exp family ori}. The gauge transformation between $\alpha'$ and $\alpha$ also induces transformation between parameters $\beta'$ and $\beta'$ in the canonical form through \eqref{Maximal Likelihood equation for Exponential Family Distributions: form transformation}.

Gauge transformations leave the exponential family model \eqref{Maximal Likelihood equation for Exponential Family Distributions: exp family ori} unchanged because $\boldsymbol{\alpha'}^T \cdot \boldsymbol{\phi'}(\bold{u}) = \boldsymbol{\alpha}^T \cdot \boldsymbol{\phi}(\bold{u})$. Therefore gauge transformations allow us to alter the moments and parameters without changing the exponential family model.

Locally, gauge transformation is equivalent to a linear coordinate transformation. However, it differs from linear coordinate transformation globally because it allows $A_{ij}$ to change with spatial position $\mathbf{x}$ and time $t$. In this case, the moment equations \eqref{The ME equation:subeq1} no longer holds for $M_i'$
\begin{equation}\label{The ME equation 1d gauge full}
     \frac{\partial M_i' }{\partial t} +  \frac{\partial F_{i \alpha}' }{\partial x_\alpha} \neq \frac{\Tilde{M}_i'-M_i'}{\tau}
\end{equation}
though the underlying physics is still the same. We could use gauge transformation to put the moment equations into locally optimized coordinate systems that are numerically more stable.

Overall, gauge transformation changes moments, parameters, and moment equations simultaneously without changing the underlying physics. As we will discuss later, it plays an important role in improving numerical stability.

\section{Simulating the Maximal Entropy
Moment Equations at Single Precision}

\subsection{Avoiding numerical overflow with the canonical form of the exponential family model}\label{subsubsec2}



This section focuses on the mapping \eqref{The ME equation:subeq2} from parameters to moments approximated with a numerical integration rule
\begin{equation} \label{Numerical integration}
\int f(\mathbf{u}) d^3\mathbf{u} \approx \sum_{p=1}^P w_p f(\mathbf{u}_p),
\end{equation}
in which $w_p$ are fixed non-negative weights and $\bold{u}_p$ are fixed nodes.

Numerical overflow is the major difficulty in implementing the mapping from parameters to moments at single precision. It refers to the overflow of the likelihood \eqref{Maximal Likelihood equation for Exponential Family Distributions: exp family ori} caused by the exponential function, whose value may exceed the upper bound of the floating-point number. Specifically, at single precision, overflow happens if the exponent $\alpha_i \phi_i(\mathbf{u})$ exceeds 80, which is reached easily for moderately large molecule velocity $\bold{u}$, especially when sufficient statistics $\{\phi_i\}$ contain high-order polynomials. An overflowed function returns `inf,' breaking down the whole simulation. It undermines the robustness of the simulation of maximal entropy moment equations at double precision and becomes a fatal issue at single precision.

We prevent numerical overflow by utilizing the canonical form \eqref{Maximal Likelihood equation for Exponential Family Distributions: exp family} of the exponential family model. The canonical form enables us to employ the log-sum-exp trick, which is widely adopted in the machine learning community preventing numerical overflow of discrete distributions. Specifically, the canonical form \eqref{Maximal Likelihood equation for Exponential Family Distributions: exp family} introduces the partition function $Z(\boldsymbol{\beta})$ to be approximated by numerical integration
\begin{equation} \label{Numerical integration partition}
Z(\boldsymbol{\beta})\approx\sum_{p=1}^P w_p \exp(\beta_i \phi_i(\mathbf{u}_p)) ; \quad i=1,\cdots,M
\end{equation}
in which we have omitted the dependencies of the natural parameters $\boldsymbol{\beta}$ on $t$ and $\bold{x}$. The partition function separates the major source of numerical overflow, the exponents in \eqref{Numerical integration partition}, from moment computation and deals with it using the log-sum-exp trick.

The moments \eqref{The ME equation:subeq2} approximated by numerical integration of the canonical form distribution are
\begin{equation} \label{Numerical integration moments}
\begin{split}
       M_{j}(t,\bold{x})&\approx \beta_0(t,\bold{x})
    \sum_{p=1}^P \phi_j(\bold{u}_p) \sigma_p(t,\bold{x})\\
\sigma_p(t,\bold{x}) &= w_p   \exp(\beta_i(t,\bold{x}) \phi_i(\mathbf{u}_p) - \log Z(\boldsymbol{\beta}) ) ; \quad i=1,\cdots,M,
\end{split}
\end{equation}
in which $\sigma_p$ is the discretized probability of molecule velocity $\bold{u}_p$. Though containing the exponential functions, this discretized probability $\sigma_p$ does not encounter numerical overflow if the partition function $Z$ is computed with the log-sum-exp trick
\begin{equation}
\label{MLE optimization goal log sum exp}
	\begin{split}
\log Z(\boldsymbol{\beta}) &\approx \log \left( \sum_{p=1}^P w_p \exp(\beta_i \phi_i(\mathbf{u}_p)) \right)\\
 &= Q + \log \left( \sum_{p=1}^P  \exp(\beta_i \phi_i(\mathbf{u}_p) + \log w_p -Q) \right); \quad  i=1,\cdots,M ,
	\end{split}
\end{equation}
in which $Q = \max_p \{\beta_i \phi_i(\mathbf{u}_p) + \log w_p \}$. Note that we have omitted the dependencies of the natural parameters $\boldsymbol{\beta}$ on $t$ and $\bold{x}$. The log-sum-exp trick prevents numerical overflow by subtracting the maximal value from all exponents. This subtraction ensures that all exponents of the exponential function are less than zero, hence eliminating the risk of numerical overflow. In summary, the log-sum-exp trick, enabled by the partition function in the canonical form, eliminated the risk of numerical overflow in computing moments using numerical integration.


It remains to determine the numerical integration method, especially the weights $w_p$ and nodes $\bold{u}_p$ in \eqref{Numerical integration} to compute the moments. In practice, numerical integration truncates the infinite integration domain on molecule velocity $\bold{u}$ to a finite integration domain, whose size is case-dependent and is determined by trial and error. On the finite integration domain, we adopt the block-wise Gauss-Legendre quadrature for numerical integration as in previous works \cite{Schaerer2017The3S, Schaerer2017EfficientAA}. Specifically, the block-wise Gauss-Legendre quadrature divides the integration domain on molecule velocity into equidistant square blocks, then integrates each block with tensor products of the one-dimensional Gauss-Legendre quadrature. Therefore the weights $w_p$ and nodes $\bold{u}_p$ for numerical integration consist of, for each block, tensor products of the weights and nodes of the one-dimensional Gauss-Legendre quadrature.

\subsection{Tackling ill-conditioned optimization with modified Newton's method}\label{subsubsec2}



Another key step in simulating the maximal entropy moment equations is to find the exponential family approximation $f_{\boldsymbol{\beta}}$ of the one-particle distribution $f$. It requires determining the map from moments $M_i$ to parameters $\boldsymbol{\beta}$. However, finding the parameters from moments is more elaborate than the direct numerical integration mapping parameters to moments in the last section. 

The common practice \cite{Alldredge2012HighOrderEC, Alldredge2013AdaptiveCO, Abramov2007AnIA, Schaerer2017The3S} determining $\boldsymbol{\beta}$ is solving parameters from an optimization problem via Newton's optimization method. We solve the parameters $\boldsymbol{\beta}$ in $f_{\boldsymbol{\beta}}$ by solving the optimization problem \eqref{MLE optimization goal}. It differs from but is equivalent to the optimization objectives \cite[Eq. 9]{Alldredge2013AdaptiveCO}\cite[Eq. 22]{Schaerer2017EfficientAA} in previous works, with the difference caused by adopting the canonical form \eqref{Maximal Likelihood equation for Exponential Family Distributions: exp family} of the exponential family model.


\begin{algorithm}[H]
\scriptsize
\caption{The Newton's Optimization Method Modified with Adding a Multiple of the Identity}\label{Newton method}
\begin{algorithmic}
\Function{NewtonOptimization}{$\bold{M}$, $\boldsymbol{\beta}$, $\boldsymbol{\phi}$}\Comment{$\bold{M}$ represents moments; $\boldsymbol{\beta}$ is the initial value of parameters; $\boldsymbol{\phi}$ represents sufficient statistics. }
\State $converged \gets $False$;\ n \gets 0;\ tol \gets 1\times 10^{-8};\  maxI \gets 400$; \ $\Delta\boldsymbol{\beta}\gets \bold{0}$;
\While{$converged$ is $False$ and $n\le maxI$ }
\State $\alpha \gets \Call{BackTrackingLineSearch}{\boldsymbol{\beta}, \Delta\boldsymbol{\beta}, \bold{M}, \boldsymbol{\phi}}$ \Comment{$\alpha$ is the step size}
\State $\boldsymbol{\beta} \gets \boldsymbol{\beta} + \alpha \Delta\boldsymbol{\beta}$ \Comment{Update the parameters}
\State $H \gets \nabla^2_{\boldsymbol{\beta}} L(\boldsymbol{\beta}; \bold{M}, \boldsymbol{\phi}$);\ $G \gets \nabla_{\boldsymbol{\beta}} L(\boldsymbol{\beta}; \bold{M}, \boldsymbol{\phi})$ \Comment{Compute the Hessian and gradient of the objective $L$ \eqref{MLE optimization goal}}
\State $\lambda \gets 1\times 10^{-3}$
\State $L_H,fail \gets \Call{CholeskyDecomposition}{H}$ \Comment{Standard Cholesky Decomposition Algorithm}
\While{$fail$} \Comment{$fail$ is $True$ if Cholesky decomposition failed}
    \State $L_H,fail \gets \Call{CholeskyDecomposition}{H + \lambda I}$ \Comment{Adding a multiple of the identity}
    \State $\lambda \gets 10\times\lambda$
\EndWhile
\State $\bold{w} =  \Call{TriangularSolve}{L_H, -G}$ \Comment{Solve $L_H \bold{w} = -G$ in which $L_H$ is lower triangular matrix}
\State $\Delta\boldsymbol{\beta} =  \Call{TriangularSolve}{L_H^T, \bold{w} }$ \Comment{Compute the update direction of the Newton's method}
\State $res \gets  0.5 \times \Delta\boldsymbol{\beta} \cdot \nabla L(\boldsymbol{\beta}) $
\State $converged \gets res \le tol$ or $\alpha \le 1\times10^{-6}$ \Comment{Check Convergence}
\State $n\gets n +1$
\EndWhile
\State \textbf{return} $\boldsymbol{\beta}$
\EndFunction
\Function{BackTrackingLineSearch}{$\boldsymbol{\beta}$, $\Delta\boldsymbol{\beta}, \bold{M}, \boldsymbol{\phi}$} \Comment{Backtracking line search to determine the step size}
\State $\alpha \gets 2$; $s \gets 0$;\ $maxT \gets 25$;\ $satisfied \gets False$
   \While{$satisfied$ is $False$ and $s \le maxT$}
   \If{not $satisfied$}
   \State $\alpha \gets 0.5 \alpha$
    \EndIf

   \State $satisfied \gets \Call{ArmijoCondition}{\boldsymbol{\beta},\alpha \Delta\boldsymbol{\beta}, \bold{M}, \boldsymbol{\phi}}$
   \State $s\gets s + 1$
   \EndWhile
\State \textbf{return} $\alpha$
\EndFunction

\Function{ArmijoCondition} {$\boldsymbol{\beta}$, $\Delta\boldsymbol{\beta}, \bold{M}, \boldsymbol{\phi}$} \Comment{The Armijo's condition as stopping criterion of line search}
\State $c\gets 5\times 10^{-4}$;\ $atol \gets 5\times 10^{-6}$;\ $rtol \gets 5\times 10^{-5}$
\State $gradD = -c \nabla_{\boldsymbol{\beta}} L(\boldsymbol{\beta}; \bold{M}, \boldsymbol{\phi})\cdot\Delta\boldsymbol{\beta}$
\State $LD = L(\boldsymbol{\beta}; \bold{M}, \boldsymbol{\phi} ) - L(\boldsymbol{\beta} + \Delta\boldsymbol{\beta};\bold{M}, \boldsymbol{\phi})$

\State \textbf{return} $(LD - gradD) \ge -( atol + rtol\times |LD| ) $ \Comment{The RHS is $0$ with floating-point error tolerance}
\EndFunction

\end{algorithmic}
\end{algorithm}



The major difficulty in solving the parameters from moments is not overflow but the ill-conditioned objective. The objective is ill-conditioned when its Hessian \eqref{MLE optimization goal hessian} is nearly singular. Since the elements of the Hessian are covariances \eqref{correlation} among sufficient statistics, the optimization objective is ill-conditioned if and only if the covariance matrix among sufficient statistics is nearly singular. Nearly singular Hessian may break optimization down, especially for Newton's optimization method utilizing curvature information in the Hessian. Specifically, Newton's method solves parameters iteratively with the optimization step
\begin{equation}
\label{Newton Optimization step}
	\begin{split}
\boldsymbol{\beta}^{(n+1)} = \boldsymbol{\beta}^{(n)} - \alpha^{(n+1)}  H^{-1} \nabla L(\boldsymbol{\beta}^{(n)}); \quad H = \nabla^2 L(\boldsymbol{\beta}^{(n)}),
	\end{split}
\end{equation}
in which $\boldsymbol{\beta}^{(n)} = \{\beta_0^{(n)},\cdots, \beta_M^{(n)}\}$, $\alpha^{(n)}$ is the step size of the $n$th optimization step, $H$ is the Hessian matrix of the optimization objective $L$ in \eqref{MLE optimization goal}. Notably, Newton's optimization relies on the inverse of the Hessian matrix, which does not exist if the Hessian is singular. In practice, computing the inverse of the Hessian fails easily, especially at single floating-point precision, if the Hessian is too ill-conditioned to be nearly singular. Failing to find the inverse of the ill-conditioned Hessian breaks down the Newton's optimization step \eqref{Newton Optimization step} hence the entire simulation.

There have been efforts to invert the ill-conditioned Hessian \cite{Abramov2007AnIA,Alldredge2013AdaptiveCO}. The general idea is converting the Hessian into the identity matrix by maintaining a stable Cholesky decomposition of the Hessian $H = L_H L_H^T$. Such decomposition involves complicated orthogonalization and iteration steps in order to decompose ill-conditioned Hessians, of which the standard decomposition fails. However, the orthogonalization and iteration steps complicate Newton's optimization method and increase the difficulty in implementation.

Alternatively, we invert the ill-conditioned Hessian by imposing a classical modification: adding a multiple of the identity \cite[p.51]{Nocedal2006}. This modification is simple but effective, despite its inaccuracy in decomposing the Hessian. Specifically, if the Hessian $H$ is ill-conditioned, we compute its inverse as
\begin{equation}
\label{Newton Optimization step}
H^{-1} = (L_H^{-1})^T L_H^{-1}; \quad
H + \lambda I = L_HL_H^T,
\end{equation}
in which $\lambda$ is a small positive number, $L_H$ is an lower-triangular matrix obtained by the Cholesky decomposition of $H + \lambda I$. The modification term $\lambda I$ removes the singularity in Hessian, enabling us to compute $L$ with the standard Cholesky decomposition. This modification is simple, easy to implement, and effectively handles ill-conditioned cases in maximal entropy moment equations at single floating-point precision.

With the Hessian modification, we could obtain the exponential family approximation $f_{\boldsymbol{\beta}}$ of $f$ by solving the optimization problem with Newton's method, even for ill-conditioned cases. The detailed implementation of Newton's optimization method is given in Algorithm \ref{Newton method}. However, the modification on Hessian becomes large if the optimization problem is too ill-conditioned. Consequently, we lose the curvature information in the Hessian, which slows the convergence of the optimization process. Moreover, the solution is inaccurate for ill-conditioned optimization, even if the optimization converges with a small residual. These issues indicate that, compared to handling the ill-conditionedness in the optimization algorithm, it is more efficient to avoid proposing an ill-conditioned optimization problem at the beginning. This could be achieved by choosing sufficient statistics according to the flow property, as we will discuss in the next section.

\subsection{Accelerating the convergence of optimization by gauge transformation}\label{subsubsec3}


Direct implementation of sufficient statistics \eqref{Sufficient 35} makes the optimization problem \eqref{MLE optimization goal} ill-conditioned even for flow near equilibrium\cite{Schaerer2017The3S}, which slows down or even breaks the Newton's method. There are attempts tackling these issues by decorrelating and normalizing sufficient statistics inside the optimization algorithm \cite{Alldredge2013AdaptiveCO,Schaerer2017The3S}, but they leave the ill-conditioned optimization problem \eqref{MLE optimization goal} untouched. Contrarily, this work aims to avoid proposing an ill-conditioned optimization problem at the beginning.

The optimization problem \eqref{MLE optimization goal} is ill-conditioned if and only if the map between parameters $\boldsymbol{\beta}$ and moments $M_i$ is ill-conditioned. It is because the Hessian of the optimization objective is closed related to the Jacobian of the map between parameters and moments
\begin{equation}
\label{moments vs parameter jacob}
\frac{\partial M_i}{\partial \beta_j} = \beta_0 H_{ij},\quad i\ge1, j \ge 1,
\end{equation}
in which $\beta_i, i\ge 1$ are the natrual parameters of the exponential family and $M_i, i\ge 1$ are moments of the sufficient statistics. This relationship indicates that selecting a proper pair of parameters and moments is the only way to avoid an ill-conditioned optimization objective, which could be achieved by gauge transformation.



This work considers gauge transformations that decorrelate sufficient statistics with translation and dilation. For 1D flow with cylindrical symmetry in velocity, we consider the gauge transformation characterized by invertible matrix $A(\mathbf{g})$ parameterized by three gauge parameters $\mathbf{g} = \{w_x,s_x,s_r\}$. Besides, we allow these parameters to have spatial and temporal variation. The gauge transformation converts the sufficient statistics $\phi_i(\bold{u})$ like \eqref{Sufficient 35} into 
\begin{equation}\label{Sufficient 35 ortho}
\begin{split}
\phi_i(\bold{u};\mathbf{g})& =\phi_i(\bar{\bold{u}}) =A_{ij}(\mathbf{g})\phi_j(\bold{u}); \quad i=0,\cdots,M\\
\bar{\bold{u}} &= \{\frac{u_x - w_x}{s_x}, \frac{u_r}{s_r}\},
\end{split}
\end{equation}
in which $u_x$, $u_r$ are molecule velocity in $x$ and radial direction. Note that $w_x$ is the translation of molecule velocity in the $x$ direction and $s_x, s_r$ represents the dilation of the molecule velocity in $x$ and radial direction. The induced transformation on moments, natural parameters, and fluxes are as follows
\begin{equation}
\label{Hermite Gauge moments}
M_i(\mathbf{g}) = A_{ij}(\mathbf{g}) M_j\quad \alpha_i(\mathbf{g}) = (A^T)^{-1}_{ij}(\mathbf{g}) \alpha_j \quad  F_{i\alpha}(\mathbf{g}) = A_{ij}(\mathbf{g}) F_{j\alpha},
\end{equation}
in which $M_i(\mathbf{g})$, $\alpha_i(\mathbf{g})$, and $F_{i\alpha}(\mathbf{g})$ are denoted as moments, parameters and fluxes  in the gauge $\mathbf{g}$.
In addition, we also define the transformation rules for moments, fluxes, and parameters in different gauges specified by gauge parameters $\bold{g} = \{w_x,s_x,s_r\}$ and $\bold{g}' = \{w_x',s_x',s_r'\}$ as follows
\begin{equation}\label{Change gauge M F}
\begin{split}
    M_i(\bold{g}') &= T_{ij}(\bold{g}',\bold{g})M_j(\bold{g})\quad F_i(\bold{g}') =T_{ij}(\bold{g}',\bold{g})F_j(\bold{g}) \\ \alpha_i(\bold{g}') &=T_{ij}^{T}(\bold{g}, \bold{g}')\alpha_j(\bold{g}) \quad  T_{ij}(\bold{g}',\bold{g}) = A_{ik}\left(\bold{g}'\right)A^{-1}_{kj}\left(\bold{g}\right),
\end{split}
\end{equation}
in which $T_{ij}$ is an analytical matrix-valued function of the gauge parameters. In practice, $T_{ij}$ is directly implemented using its analytical form rather than computed from the matrix $A$ to avoid the computational demanding matrix inversion. 

Putting these all together, we have defined gauge transformation \eqref{Sufficient 35 ortho} as the translation, dilation, and decorrelation of sufficient statistics specified by gauge parameters $\bold{g} = \{w_x,s_x,s_r\}$, which also transform moments, fluxes, and natural parameters into the gauges $G$ as in \eqref{Hermite Gauge moments}. Different gauge parameters transform moments, fluxes, and natural parameters into different gauges. But they are physically equivalent representations of the same exponential family distribution related by \eqref{Change gauge M F}.

We focus on two gauges of interest: The trivial gauge and the Hermite gauge. The trivial gauge is specified by the gauge parameter $\mathbf{g}_0 = \{0,1,1\}$, indicating that no translation and dilation is performed on the molecule velocity. The Hermite gauge is specified by the gauge parameter $\mathbf{g}_H = \{w_{H,x},s_{H,x},s_{H,r}\}$ such that 
\begin{equation} \label{The Hermite Gauge moments}
M_1(\mathbf{g}_H) = M_2(\mathbf{g}_H) = M_3(\mathbf{g}_H) = 0.
\end{equation}
These constraints on moments specify the following gauge parameters
\begin{equation} \label{Hermite parameters}
\begin{split}
    w_{H,x} &= \frac{1}{n}\int u_x f_{\boldsymbol{\beta}}(\bold{u}) d^3 \bold{u}; \\ s_{H,x}^2 &= \frac{1}{n}\int (u_x-w_{H,x})^2 f_{\boldsymbol{\beta}}(\bold{u}) d^3 \bold{u}\\ s_{H,r}^2 &= \frac{1}{2n}\int( u_y^2+u_z^2 )f_{\boldsymbol{\beta}}(\bold{u}) d^3 \bold{u},
\end{split}
\end{equation}
in which $n = M_0$ is the number density, $f_{\boldsymbol{\beta}}$ is the exponential family approximation \eqref{Maximal Likelihood equation for Exponential Family Distributions: exp family ori} of the one-particle distribution $f$. 

In practice, gauge parameters of the Hermite gauge are computed analytically from moments $\mathbf{M}(\mathbf{g})$ in an arbitrary gauge $\mathbf{g}$ as follows
\begin{equation} \label{Hermite parameters in moments}
\begin{split}
    \bold{g}_H = H(\mathbf{M}(\mathbf{g}),\mathbf{g} )\cdot \mathbf{M}(\mathbf{g})
\end{split}
\end{equation}
in which $H(\mathbf{M}(\mathbf{g}),\mathbf{g} )$ is a matrix of size $3\times 9$ whose components are analytical functions of moments $\mathbf{M}(\mathbf{g})$ and gauge parameter $\mathbf{g}$. As we will see in Section \ref{result acc}, transforming moments and parameters into the Hermite gauge significantly accelerates Newton's method in solving the optimization problem \eqref{Maximal Likelihood equation for Exponential Family Distributions: mle exp family}.


\subsection{Integrating Gauge Transformation into Finite Volume PDE Solver}\label{subsubsec3}

In this section, we solve the one-dimensional maximal entropy moment equations utilizing a finite volume PDE solver. The maximal entropy moment equations form a hyperbolic system of conservation laws with source terms, which could be handled by a finite-volume PDE solver. However, we need to integrate gauge transformation of moments and fluxes into the finite volume solver to keep the maximal entropy moment equations well-conditioned, which is the focus of this section.


The major difficulty introduced by gauge transformation is the temporal and spatial variation of gauge parameters. The maximal entropy moment equations \eqref{The ME equation:subeq1} are no longer valid when the gauge parameters $\bold{g}$ have temporal and spatial variation 
\begin{equation}\label{The ME equation 1d gauge full}
     \frac{\partial M_i(t, \boldsymbol{x}, \bold{g}) }{\partial t} +  \frac{\partial F_{i \alpha}(t, \boldsymbol{x}, \bold{g}) }{\partial x_\alpha} \neq \frac{\Tilde{M}_i(t, \boldsymbol{x}, \bold{g})-M_i(t, \boldsymbol{x}, \bold{g})}{\tau}.
\end{equation}
A possible way to address this issue is to include the dynamics of gauge parameters in the moment equations. However, this would make the moment equations very complicated. Therefore, we choose a simpler but \textbf{exact} solution: keeping the gauge parameters locally constant whenever derivatives are calculated.

Temporal changes in the gauge parameter field could be handled easily by keeping the gauge parameters locally constant. Specifically, we split each time step into two separate steps
\begin{equation} \label{The ME Equation time split}
\begin{split}
    &\mbox{1. Advance the time: $M(t+\Delta t, \bold{x}, g(t, \bold{x})) \xleftarrow{} M(t, \bold{x}, g(t, \bold{x}))$ } \\
    &\mbox{2. Gauge transformation: $M(t+\Delta t, \bold{x}, g(t+\Delta t, \bold{x})) \xleftarrow{} M(t+\Delta t, \bold{x}, g(t, \bold{x}))$}.\\
\end{split}
\end{equation}
This splitting eliminated the time dependency of gauge parameters $\boldsymbol{g}$ when updating the moments, allowing us to use the original updating rule of maximal entropy moment equations.

The spatial variation of the gauge parameter field requires more careful treatment in the flux when using finite volume solvers, which we will discuss step by step. 

First, we consider the 1D maximal entropy moment equations in the gauge $\bold{g}$ specified by fixed gauge parameters $\bold{g} = \{w_x, s_x, s_r\}$ in \eqref{Hermite parameters in moments} without the source term
\begin{equation}\label{The ME equation 1d H}
     \frac{\partial M_i(t, x;\bold{g}) }{\partial t} +  \frac{\partial F_{i}(t, x;\bold{g}) }{\partial x} = 0;\quad i=0,\cdots,M,
\end{equation}
in which $M_i(\bold{g})$, $F_i(\bold{g})$ are moments and fluxes in the gauge $\bold{g}$ defined in \eqref{Hermite Gauge moments}. These moment equations are hyperbolic conservation laws that could be handled by finite volume solvers such as the local Lax-Friedrichs scheme. Specifically, a numerical solver discretizes the time into steps of the size $\Delta t$ and discretizes the spatial computation domain into $N$ cells with cell width $\Delta x$. It further discretizes the moment equations into
\begin{equation}\label{The ME equation fvm}
    M_i(t+\Delta t, n;\bold{g})= M_i(t, n;\bold{g})+ \frac{\Delta t}{\Delta x} \left[ F_i\left(t+\frac{\Delta t}{2}, n-\frac{1}{2};\bold{g}\right) - F_i\left(t+\frac{\Delta t}{2}, n+\frac{1}{2};\bold{g}\right) \right],
\end{equation}
in which $M_i(t, n;\bold{g})$ is moments in the $n$th cell at time $t$ at constant gauge parameters $\bold{g} = \{w_x, s_x, s_r\}$, and $F_i(t, n\pm\frac{1}{2};\bold{g})$ is fluxes at the $n\pm\frac{1}{2}$th cell interface $t$ at constant gauge parameters $\bold{g} = \{w_x, s_x, s_r\}$. The fluxes $F_i$ at time $t+\frac{\Delta t}{2}$ at cell interfaces $n\pm\frac{1}{2}$ in \eqref{The ME equation fvm} could be any numerical flux such as the first-order local Lax-Friedrichs scheme
\begin{equation}\label{The lax F}
\begin{split}
    F_i^{LF}\left(t+\frac{\Delta t}{2}, n-\frac{1}{2};\bold{g}\right)  &= \frac{1}{2}\left( F_i\left(t, n-1;\bold{g}\right)  + F_i\left(t, n;\bold{g}\right)  \right) \\& \quad + \frac{\lambda_m}{2}\left( M_i(t, n-1;\bold{g}) - M_i(t, n;\bold{g})  \right)\\
    \lambda_m &= \max_p\{ |\lambda_{p}(n-1)|, |\lambda_{p}(n)| \} \\\lambda_{p}(n) &= p\mbox{th eignvalue of } \frac{\partial F(\bold{g})}{\partial M(\bold{g})},
\end{split}
\end{equation}
the second-order Lax-Wendroff scheme
\begin{equation}\label{The lax W}
\begin{split}
   M_i(t, n-\frac{1}{2};\bold{g}) &= \frac{1}{2} \left( M_i(t, n-1;\bold{g}) + M_i(t, n;\bold{g}) \right) \\& \quad + \frac{1}{2} \frac{\Delta t}{\Delta x}  \left( F_i\left(t, n-1;\bold{g}\right)  - F_i\left(t, n;\bold{g}\right) \right) \\
   \beta_i(t, n-\frac{1}{2};\bold{g}) &= \mbox{argmin}_\beta \ L( \boldsymbol{\beta}; \bold{M}(t, n-\frac{1}{2};\bold{g}), \boldsymbol{\phi}) \\
    F_i^{LW}\left(t+\frac{\Delta t}{2}, n-\frac{1}{2};\bold{g}\right)  &= F_i\left( \beta_i(t, n-\frac{1}{2};\bold{g}); \mathbf{g} \right)\\
\end{split}
\end{equation}
in which the $F_i\left( \beta_i(t, n-\frac{1}{2};\bold{g}); \mathbf{g} \right)$ is computed according to \eqref{The ME equation:subeq3}, or the second-order flux limiter centered (FLIC) scheme
\begin{equation}\label{The FLIC}
\begin{split}
    F_i^{FLIC}\left(t+\frac{\Delta t}{2}, n-\frac{1}{2};\bold{g}\right)  &= F_i^{LF}\left(t+\frac{\Delta t}{2}, n-\frac{1}{2};\bold{g}\right)  \\ + \phi_{n-\frac{1}{2}} &\left( F_i^{LW}\left(t+\frac{\Delta t}{2}, n-\frac{1}{2};\bold{g}\right) -F_i^{LF}\left(t+\frac{\Delta t}{2}, n-\frac{1}{2};\bold{g}\right)  \right)\\
    \phi_{n-\frac{1}{2}} &= \frac{r_{n-\frac{1}{2}}+|r_{n-\frac{1}{2}}|}{1+r_{n-\frac{1}{2}}}\\ r_{n-\frac{1}{2}} &= \frac{M_i(t,n-1;\mathbf{g}) - M_i(t,n-2;\mathbf{g}) }{ M_i(t,n;\mathbf{g}) - M_i(t,n-1;\mathbf{g})  },
\end{split}
\end{equation}
in which we use the Van-Leer flux limiter $\phi$ to ensure total variation diminishing (TVD).

Second, we modify the above discretization, allowing the gauge parameters $\bold{g}$ to have spatial variation. We handle the spatial variation of gauge parameters by allowing the $n$th cell to possess its gauge parameters $\bold{g}(n) = \{w_x(n),s_x(n),s_r(n)\}$. In this case, the discretized moments equation \eqref{The ME equation fvm} is still valid if all its quantities use the same gauge parameters $\bold{g}(n)$.
\begin{equation}\label{The ME equation fvm 2}
\begin{split}
    M_i(t+\Delta t, n;\bold{g}(n))&= M_i(t, n;\bold{g}(n))+ \\ \frac{\Delta t}{\Delta x}& \left[ F_i\left(t+\frac{\Delta t}{2}, n-\frac{1}{2};\bold{g}(n)\right) - F_i\left(t+\frac{\Delta t}{2}, n+\frac{1}{2};\bold{g}(n)\right) \right],
\end{split}
\end{equation}
in which the fluxes $F_i$ are computed at the gauge $\mathbf{g}(n)$.
The key difference between constant gauge parameter and those with spatial variation is that the latter involves gauge transformation of moment and fluxes in the gauge $\bold{g}(n-1)$, such as $M_i\left(t, n-1;\bold{g}(n-1)\right)$ and $F_i\left(t, n-1;\bold{g}(n-1)\right)$, into the gauge $\mathbf{g}(n)$ before computing fluxes $F_i\left(t+\frac{\Delta t}{2}, n\pm\frac{1}{2};\bold{g}(n)\right)$. In practice, these moments and fluxes are not stored in memory but are computed from $M_i\left(t, n-1;\bold{g}(n-1)\right)$ and $F_i\left(t, n-1;\bold{g}(n-1)\right)$ by gauge transformations between neighboring cells
\begin{equation}\label{The lax F convert}
\begin{split}
    M_i\left(t, n\pm1;\bold{g}(n)\right) &= T_{ij}\left(\bold{g}(n),\bold{g}(n\pm1)\right) M_j\left(t, n\pm1;\bold{g}(n\pm1)\right)\\
    F_i\left(t, n\pm1;\bold{g}(n)\right) &= T_{ij}\left(\bold{g}(n),\bold{g}(n\pm1)\right) F_j\left(t, n\pm1;\bold{g}(n\pm1)\right)
\end{split}
\end{equation}
in which $T_{ij}$ are defined in \eqref{Change gauge M F}. Therefore, spatial variation of gauge parameters is incorporated in the numerical solver by transforming neighboring cells into the same local gauge. In summary, the discretization of moment equations without source term consists of \eqref{The lax F}, \eqref{The lax F convert}, and \eqref{The ME equation fvm 2}, in which \eqref{The lax F} and \eqref{The ME equation fvm 2} implement the local Lax-Friedrichs scheme while \eqref{The lax F convert} handle the spatial variation of gauge parameters by corresponding gauge transformation.

To this end, we deal with the source term in the RHS of moment equations \eqref{The ME equation} by operator splitting. For each time step from $t$ to $t+\Delta t$, we solve the following equation in addition to the moment equations \eqref{The ME equation 1d H} without source 
\begin{equation}\label{The ME equation 1d H source}
     \frac{\partial M_i(\bold{g}) }{\partial t}  = \frac{\Tilde{M}_i(\bold{g})-M_i(\bold{g})}{\tau}\quad \quad i=0,\cdots,M,
\end{equation}
in which $\bold{g} = \{w_x, s_x, s_r\}$, $\Tilde{M}_i(\bold{g}) = A_{ij}(\bold{g})\Tilde{M}_i$ are moments \eqref{The ME equation:subeq4} of the Maxwell distribution \eqref{Maximal Likelihood equation for Exponential Family Distributions: The Maxwell distribution} in the gauge $\bold{g}$ defined in \eqref{Hermite Gauge moments}. In practice, we update the moments according to the following update rule
\begin{equation}\label{source update}
\begin{split}
    M_i\left(t, n;\bold{g}(n)\right) \xleftarrow{} \left(1-\frac{\Delta t}{\tau}\right)M_i\left(t, n;\bold{g}(n)\right) + \frac{\Delta t}{\tau}  \Tilde{M}_i\left(t, n;\bold{g}(n)\right)
\end{split}
\end{equation}
in which $\xleftarrow{}$ indicates substitute $M_i\left(t, n;\bold{g}(t, n)\right)$ with the expression in the RHS of \eqref{source update}. This update rule is applied after each time step described by \eqref{The ME equation fvm 2}.

We summarize the discretization of the moment equations in Alg.\ref{TimeStep} as the local scheme coupled with gauge transformation to mitigate ill-conditionedness. In practice, we implement the numerical solving using the JAX \cite{jax2018github} library of Python, which admits on both CPU and GPU backend. Its auto-differentiation function allows us to evaluate gradients and Hessians easily for Newton's optimization method. 

\begin{algorithm}[H]
\scriptsize
\caption{One Time Step in Simulating the Maximal Entropy Moment Equations with Gauge Transformation}\label{TimeStep}
\begin{algorithmic}
\Function{Transport}{$\bold{M}, \boldsymbol{\beta},\bold{g}, \Delta t$} \Comment{Input moments, natural parameters, and gauge parameters for all cells}
\State $\boldsymbol{\beta}\gets \Call{NewtonOptimization}{\bold{M},\boldsymbol{\beta}, \boldsymbol{\phi}(\bold{u} ;\mathbf{g})}$ \Comment{Solve the natural parameters by Alg. \ref{Newton method} in the gauge $\mathbf{g}$}
\State $\bold{F} \gets \Call{ComputeFluxes}{\boldsymbol{\beta}, \boldsymbol{\phi}(\bold{u} ;\mathbf{g})}$\Comment{Compute fluxes of the statistics $\boldsymbol{\phi}(\bold{u} ;\mathbf{g})$ by \eqref{The ME equation:subeq3} }
\State $\bold{M}_{\pm1}, \bold{F}_{\pm1} \gets \Call{SpatialGaugeTransformation}{\bold{M}, \bold{F}, \bold{g}}$\Comment{Perform gauge transformation as in \eqref{The lax F convert} }
\State $\bold{M} \gets \Call{FiniteVolumeStep}{\bold{M}, \bold{F},\bold{M}_{\pm1}, \bold{F}_{\pm1}, \bold{g}, \Delta t}$\Comment{One step forward in time by \eqref{The ME equation fvm 2} }
\State \textbf{return} $\bold{M}, \boldsymbol{\beta}, \bold{g}$
\EndFunction

\Function{Collision}{$\bold{M}, \boldsymbol{\beta},\bold{g}, \Delta t$} \Comment{Input moments, natural parameters, and gauge parameters for all cells}
\State $\bold{M} \gets \Call{SourceTerm}{\bold{M},  \bold{g}, \Delta t}$\Comment{Compute source term by operator splitting \eqref{source update} }
\State \textbf{return} $\bold{M}, \boldsymbol{\beta}, \bold{g}$
\EndFunction

\Function{Step}{$\bold{M}, \boldsymbol{\beta},\bold{g}$} \Comment{Input moments, natural parameters, and gauge parameters for all cells}

\State $\bold{M}, \boldsymbol{\beta}, \bold{g} \gets \Call{Collision}{\bold{M},\boldsymbol{\beta},\bold{g}, \Delta t/2}$\Comment{Compute collision term by operator splitting}
\State $\bold{M}, \boldsymbol{\beta}, \bold{g} \gets \Call{Transport}{\bold{M},\boldsymbol{\beta},\bold{g}, \Delta t}$\Comment{Compute Transport term by operator splitting}
\State $\bold{M}, \boldsymbol{\beta}, \bold{g} \gets \Call{Collision}{\bold{M},\boldsymbol{\beta},\bold{g}, \Delta t/2}$\Comment{Compute collision term by operator splitting}

\State $\bold{g}_H \gets \Call{ComputeGaugeParameters}{\bold{M}, \bold{g}}$\Comment{Compute the Hermite gauge parameters by \eqref{Hermite parameters in moments} }
\State $\bold{M}_H,\boldsymbol{\beta}_H \gets \Call{GaugeTransformation}{\bold{M}, \boldsymbol{\beta}, \bold{g}_H, \bold{g}}$\Comment{Transform into the Hermite gauge using \eqref{Change gauge M F} and \eqref{Maximal Likelihood equation for Exponential Family Distributions: form transformation}}
\State \textbf{return} $\bold{M}_H, \boldsymbol{\beta}_H, \bold{g}_H$
\EndFunction

\end{algorithmic}
\end{algorithm}

\subsection{The Resolution Limits of Maximal Entropy Moment Equations}\label{Resolution Limits sec}

The maximal entropy moment equations have resolution limits in spatial and temporal resolution, which we will explain from two perspectives: the invariant-manifold assumption and solvability. 

\subsubsection{Invariant manifold and the mean free path resolution limit } \label{Invariant manifold resolution}

The mean free path sets spatial resolution limits of MEM, beyond which the Boltzmann equation cannot be approximated more accurately. This resolution limits stems from the invariant-manifold assumption, which allow us to substitute $f$ with $f_{\boldsymbol{\beta}}$ in the Boltzmann equation to get the maximal entropy moment equations. We demonstrate this in the following discussion by comparing the Lie operator splitting discretizations of the maximal entropy moment equations and the Boltzmann equation.

The Lie operator splitting for maximal entropy moment equations has each time step consists the following two sub-steps\begin{equation} \label{The ME Equation split}
\begin{split}
    &\mbox{1. Compute $f_{\boldsymbol{\beta}}$ by moment matching \eqref{Maximal Likelihood equation for Exponential Family Distributions: exp family moments matching} with $f$} \\
    &\mbox{2. $f = f_{\boldsymbol{\beta}} +  \left(- u_\alpha \frac{\partial f_{\boldsymbol{\beta}}}{\partial x_\alpha} + Q(f_{\boldsymbol{\beta}},f_{\boldsymbol{\beta}}) \right)\Delta t$},\\
\end{split}
\end{equation} in which the second step solves the Boltzmann equation by the forward Euler method. This splitting is reasonable because the maximal entropy moment equations are obtained by taking moments on the Boltzmann equation with $f$ substituted by $f_{\boldsymbol{\beta}}$. 
Such "substituting and computing moments" process corresponds to the first step in \eqref{The ME Equation split} and is split from the Boltzmann equation. The splitting and the forward Euler method in \eqref{The ME Equation split} brings a global truncation error of the order $O(\Delta t)$, solving the maximal entropy moment equations at first-order accuracy.

The Lie operator splitting splits the Boltzmann equation into two sub-equations
\begin{equation} \label{The Boltzmann Equation split Euler}
\begin{split}
    &\mbox{1. $f^{(1)}$ $\xleftarrow{}$ Solve $\frac{\partial f}{\partial t} = Q(f-f_{\boldsymbol{\beta}},f - f_{\boldsymbol{\beta}}) + 2 Q(f-f_{\boldsymbol{\beta}}, f_{\boldsymbol{\beta}}) $ }  \\
    &\mbox{  \ \ \ \  to time $\Delta t$ with initial condition $f$ at $t=0$  }  \\
   &\mbox{2. $f  = f^{(1)} + \left(- u_\alpha \frac{\partial f^{(1)}}{\partial x_\alpha} + Q(f^{(1)}_{\boldsymbol{\beta}},f^{(1)}_{\boldsymbol{\beta}}) \right) \Delta t$},\\
\end{split}
\end{equation}
according to \eqref{Collision decompose}, in which $f_{\boldsymbol{\beta}}$ and $f^{(1)}_{\boldsymbol{\beta}}$ are exponential family approximations of $f$ and $f^{(1)}$. The first step in \eqref{The Boltzmann Equation split Euler} describes the relaxation from $f$ to its steady state $f_{\boldsymbol{\beta}}$, while the second step describes the transport and collision of molecules and is discretized with the forward Euler method. This operator splitting of the Boltzmann equation has a first-order accuracy, and the global truncation error is of the order $O(\Delta t)$.


The operator splitting for the Boltzmann equation \eqref{The Boltzmann Equation split Euler}  and for the maximal entropy moment equations \eqref{The ME Equation split} differ only in the first step. Instead of substitute $f$ with $f_{\boldsymbol{\beta}}$, the first step of the Boltzmann equation splitting describes the relaxation from $f$ to $f_{\boldsymbol{\beta}}$ with a finite relaxation time which we denote as $\tau_e$. Notably, $f^{(1)} = f_{\boldsymbol{\beta}}$ could be established provided the time step $\Delta t$ is no less than the time $\tau_e$. Under the condition $\Delta t \ge \tau_e$, the operator splitting \eqref{The ME Equation split} for maximal entropy moment equations is the same as \eqref{The Boltzmann Equation split Euler} hence solving the Boltzmann equation. When $\Delta t < \tau_e$, the discrepancy between the maximal entropy moment equation and the Boltzmann equation appears as a global truncation error of the order $O(\tau_e)$. We can get rid of the operator splitting error by setting the time step $\Delta t$ to zero. This reveals that the maximal entropy moment equation is a first-order approximation of the Boltzmann equation with an intrinsic error of the order $O(\tau_e)$.

The intrinsic error of order $O(\tau_e)$ is a measure of how well the maximal entropy moment equation approximates the Boltzmann equation. It depends on the relaxation time $\tau_e$, which is independent of numerical resolution. Therefore, using a very small time step $\Delta t$ that is much smaller than the relaxation time $\tau_e$ does not improve the accuracy of the numerical scheme. This means that the relaxation time $\tau_e$ sets a lower limit for the time step size $\Delta t$.


The limit for time step also leads to a lower limit $l_e$ for the spatial mesh spacing $\Delta x$. This is because the spatial mesh spacing $\Delta x$ and the time step $\Delta t$ are proportional to each other according to the CFL condition $\Delta x \ge \frac{u\Delta t}{C} $, in which $C$ is the Courant number and $u$ is the speed of propagation of the flow signal. This tells us that the mesh spacing limit $l_e = \frac{u\tau_e}{C} $ is proportional to the time step limit $\tau_e$.

The mean free path of the gas provides a useful estimation to the spatial spacing limit $l_e$:
\begin{equation}\label{limit1}
    l_e \ge \sqrt{\frac{ \pi}{2 }}\frac{\mu}{n  \sqrt{m k_B T}}	,
\end{equation}
in which $n$ is the number density, and $\mu$ is viscosity coefficient, $k_B$ is the Boltzmann coefficient, and $T$ is the temperature of gas. This is because the spacing limit aims to ensure that $f$ has been fully relaxed to its maximal entropy correspondence $f_{\boldsymbol{\beta}}$. Such relaxation is accomplished by collisions among gas particles, which generate entropy and drive the gas towards a maximum entropy state. Therefore, there must be enough collisions occurring within each cell. This means the cells in the spatial mesh should have a size at least the mean free path, since it is the average distance that a gas particle travels before colliding with another particle.

Similar to the case of time step, setting the cell size $\Delta x$ finer than $l_e$ does not provides extra accuracy in approximating the Boltzmann equation because of the intrinsic approximation error. In practice, setting $\Delta x$ several times larger than the mean free path is enough to provide accurate simulation results according to our experiments in Section \ref{continuum shock}.

\subsubsection{Solvability and the resolution limit } \label{Realizability and resolution limit}

The maximal entropy moment equations describing gases with several statistical moments. Such a description fails when we have enough resolution to distinguish single gas particles, because there are no statistics of one gas particle. Consequently, the number of gas particles inside a mesh cell sets a spatial resolution limit $l_\beta$, beyond which the optimization problem \eqref{MLE optimization goal} is too ill-conditioned to be solvable.


The particle nature of gas tells us that moments $M_i$ are sample means of $\phi_i(\mathbf{u}), i=1,\cdots, M$, given $N$ gas particles with velocity $\mathbf{u}_a, a=1, \cdots, N$:
\begin{equation}
    M_i = \frac{1}{N} \sum_{a=1}^N \phi_i (\mathbf{u}_a); \quad i=1,\cdots, M
\end{equation}
It is known that for polynomial $\phi(\mathbf{u})$, the solution of the optimization problem \eqref{MLE optimization goal} almost surely exists \cite{Crain1976ExponentialMM}, as long as the invariant manifold assumption holds and $N > M$. Therefore we know that to ensure the existence of a solution, each mesh cell should contain at least $M$ particles to make the maximal entropy optimization solvable. This tells us that the spatial resolution limit $l_\beta$ must be larger than the mean spacing of gas particles:
\begin{equation}\label{limit2}
    l_\beta > \left(\frac{M}{n}\right)^{1/3},
\end{equation}
where $M$ is the number of moments and $n$ is the number density of gas. Beyond this spatial limit, the optimization problem \eqref{MLE optimization goal} becomes too ill-conditioned to be solvable. 

In general, the optimization problem \eqref{MLE optimization goal} becomes more ill-conditioned as the spatial resolution increases. Our experiment in Section \ref{Convergence and Stability} confirms this by showing that the condition number of the Jacobian matrix \eqref{moments vs parameter jacob} increases as the mesh spacing shrinks. It also indicates that the numerical stability of the maximal entropy moment equations deteriorates when the mesh is over-refined.

It worth noting that the resolution limit \eqref{limit2} is automatically satisfied if \eqref{limit1} is satisfied, because the mean spacing of gas particles are usually much smaller than mean free path. In practice, we only need to choose a resolution respecting the spatial limit \eqref{limit1} to achieve the balance among accuracy, stability, and computational efficiency.

In summary, Section \ref{Invariant manifold resolution} and Section \ref{Realizability and resolution limit} have demonstrated the existence of lower limits of the time step and spatial mesh spacing for the maximal entropy moment equations. Though numerical convergence could be achieved by refining spatial mesh beyond the resolution limit, there is no need to reach such convergence because it does not provide extra accuracy approximating the Boltzmann equation. Moreover, an over-refined spatial mesh undermines the numerical stability of maximal entropy moment equations.

\begin{figure}[!t]
\centering
\includegraphics[width=12cm]{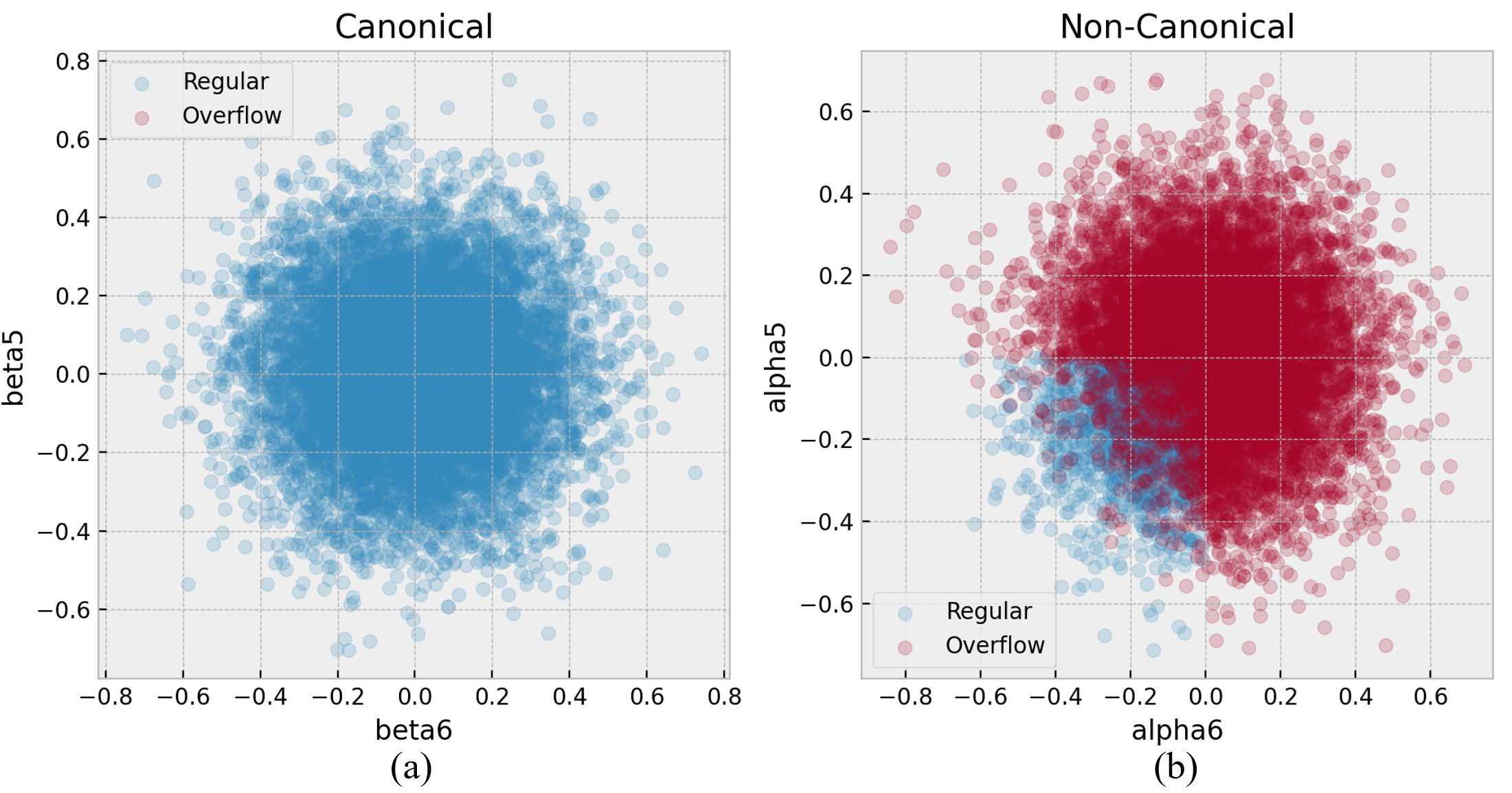}
\caption{Scatter plots show randomly chosen parameters close to an equilibrium state in the canonical and non-canonical forms, with red and blue dots indicating whether numerical overflow happens when computing moments from these parameters at single precision. (a) The scatter plot of $\beta_5$ and $\beta_6$, components of the randomly chosen parameters in the canonical form defined in \eqref{Maximal Likelihood equation for Exponential Family Distributions: exp family}. None of them encounter numerical overflow in computing moments. (b) The scatter plot of $\alpha_5$ and $\alpha_6$, components of the randomly chosen parameters in the non-canonical form defined in \eqref{Maximal Likelihood equation for Exponential Family Distributions: exp family ori}. Many encounter numerical overflow except for a small portion in the third quadrant. These two plots show that the canonical form performs better than the non-canonical form in preventing numerical overflow in the neighborhood of the equilibrium. }\label{Fig1}
\end{figure}

\section{Results}

\subsection{ The canonical form prevents numerical overflow }

This section demonstrates that the canonical form prevents numerical overflow. We examine the occurrence of numerical overflow in both the canonical and non-canonical forms when computing moments at single precision from randomly chosen parameters close to an equilibrium state. These parameters are sampled from a multivariate Gaussian distribution with a standard deviation of 0.1. For canonical and non-canonical forms, the center of this distribution are different parameters representing the same equilibrium state characterized by $n = 1$, $v_x = 0$, and $k_B T/m = 1$. In other words, we randomly pick parameters from a ball with a radius of 0.1, centered at the equilibrium state. The shape of the ball is different in the canonical and non-canonical forms due to the variation in parameterization. This difference in shape is the key factor that makes the canonical form effective in avoiding numerical overflow.

Fig.\ref{Fig1} demonstrates whether numerical overflow happens when computing moments from these randomly sampled parameters using scatter plots. Specifically, we compute moments of the statistics \eqref{Sufficient 35 ortho} for the gauge parameters $\bold{g} = \{1,1,0\}$ from randomly sampled parameters by integrating on the finite integration domain $ (u_x, u_r)$ $\in$ $ [-30.7 , 37.3] \times [0.0, 34.0] $. The integration domain is divided into $16\times16$ uniform blocks, and each of them is integrated using an 8th-order Gauss-Legendre quadrature. During the integration, any occurrence of $NaN$ or $Inf$ is recorded as numerical overflow. Fig.\ref{Fig1} uses red and blue dots to distinguish between cases where overflow occurs and cases where it does not.

Fig.\ref{Fig1}(a) is the scatter plot of $\beta_5$ and $\beta_6$ that are components of the randomly chosen parameters in the canonical form. It shows that no numerical overflow occurs during the calculation of moments. This result means that the equilibrium state has a benign neighborhood without the risk of numerical overflow in the canonical form. Fig.\ref{Fig1}(b) is the scatter plot of $\alpha_5$ and $\alpha_6$ that are components of the randomly chosen parameters in the non-canonical form. It shows that many parameters encounter numerical overflow except for a small portion in the third quadrant, where both parameters for $u_x^4$ and $u_r^4$ are strongly negative. Outside the third quadrant, one of the coefficients of $u_x^4$ and $u_r^4$ becomes positive, resulting an ill-conditioned density distribution of the form $exp(u^4)$ which blows up at the infinity. These scatter plots show the strong stability of our canonical form, which is capable of computing moments under such ill-conditioned cases hence reducing the occurrence of numerical overflow.

\begin{figure}[!t]
\centering
\includegraphics[width=6.5cm]{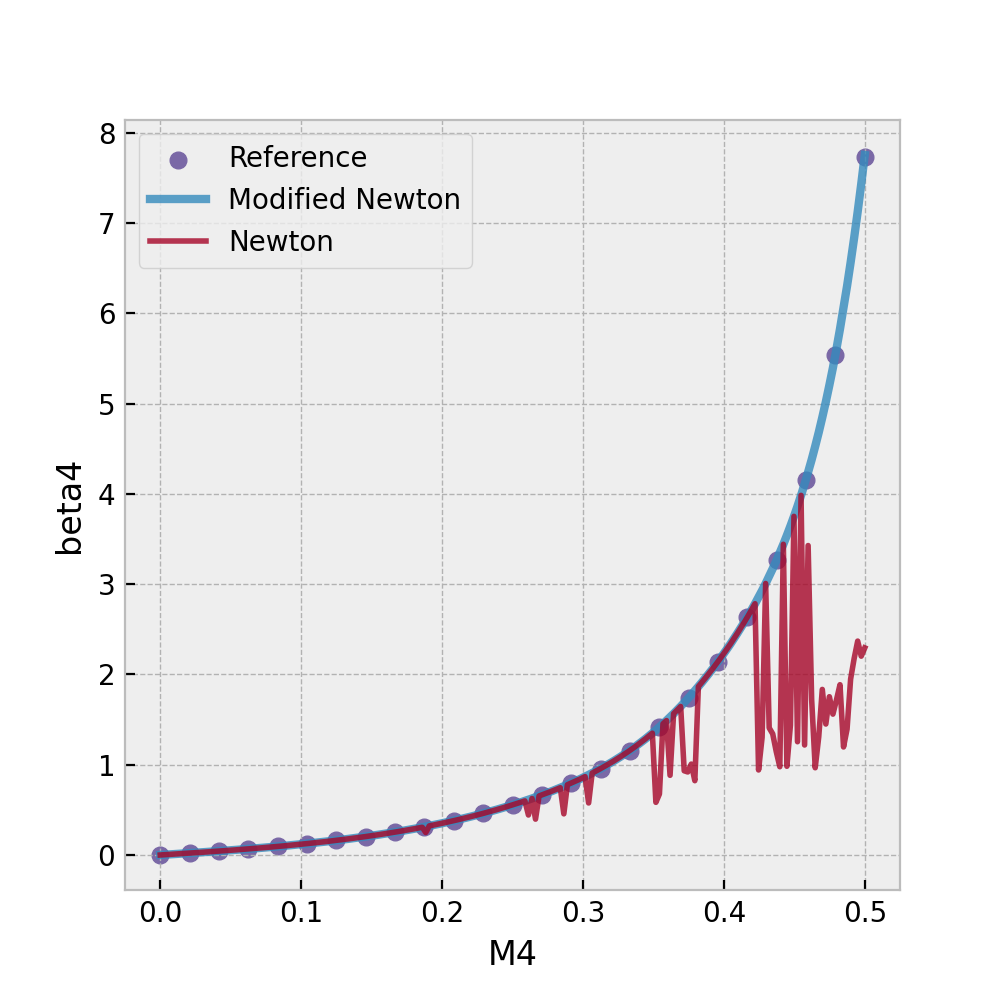}
\caption{This plot compares the performance of Newton's method and the modified Newton's method in optimizing the parameters corresponding to given moments at single precision. The parameter $\beta_4$ is plotted against the corresponding moment $M_4$ while keeping all other moments as constants. The blue line represents the parameter $\beta_4$ optimized using the modified Newton's method, which is in perfect agreement with the reference values computed at double precision shown as dots on the plot. On the other hand, the red line, representing the parameter $\beta_4$ optimized using Newton's method, fails to match the reference values beyond $M_4 = 4.5$. Overall, this plot demonstrated that the modified Newton's method is more accurate and robust in solving parameters from moments compared to the Newton's method.}\label{Fig2}
\end{figure}

\subsection{ The modified Newton's method is more accurate and robust in solving parameters } \label{result acc}

This section compares the performance of Newton's method and the modified Newton's method in optimizing parameters from moments according to the objects defined in equation \eqref{MLE optimization goal} at single precision. The methods are evaluated on a series of moments that increase from an equilibrium state value of $M_4=0$ to a non-equilibrium value of $M_4=0.5$, while all other moments remain constant. These moments correspond to the statistics given in equation \eqref{Sufficient 35 ortho} for gauge parameters $\bold{g} = \{0,1,1\}$, and the equilibrium state is characterized by $n = 1$, $v_x = 0$, and $k_B T/m = 1$. The integration in the objective is performed over the finite domain $(u_x, u_r) \in [-30.7, 37.3] \times [0.0, 34.0]$, which is divided into $16 \times 16$ uniform blocks and integrated using 8th-order Gauss-Legendre quadrature. The optimization stops when it is converged at which the residual is less than $10^{-8}$, or it exceeds the maximal iteration limit of 500.

In Fig.\ref{Fig2}, the parameter $\beta_4$ is plotted against the corresponding moment $M_4$ that ranges from an equilibrium state value of 0 to a non-equilibrium value of 0.5 while keeping all other moments as constants. The blue line represents the parameter $\beta_4$ optimized using the modified Newton's method, which perfectly matches the reference values computed at double precision shown as dots on the plot. The red line, representing the parameter $\beta_4$ optimized using Newton's method, begins to fluctuate around $M_4 = 0.3$ and fails to match the reference values beyond $M_4 = 4.5$. It is because the Hessian of the optimization objective becomes ill-conditioned as $M_4$ grows, which produces an inaccurate optimization direction. Such direction only admits extremely small step sizes, preventing Newton's method's convergence. Overall, Fig.\ref{Fig2} demonstrates that the modified Newton's method is more accurate and robust in solving parameters from moments than Newton's method.

\begin{figure}[!t]
\centering
\includegraphics[width=13cm]{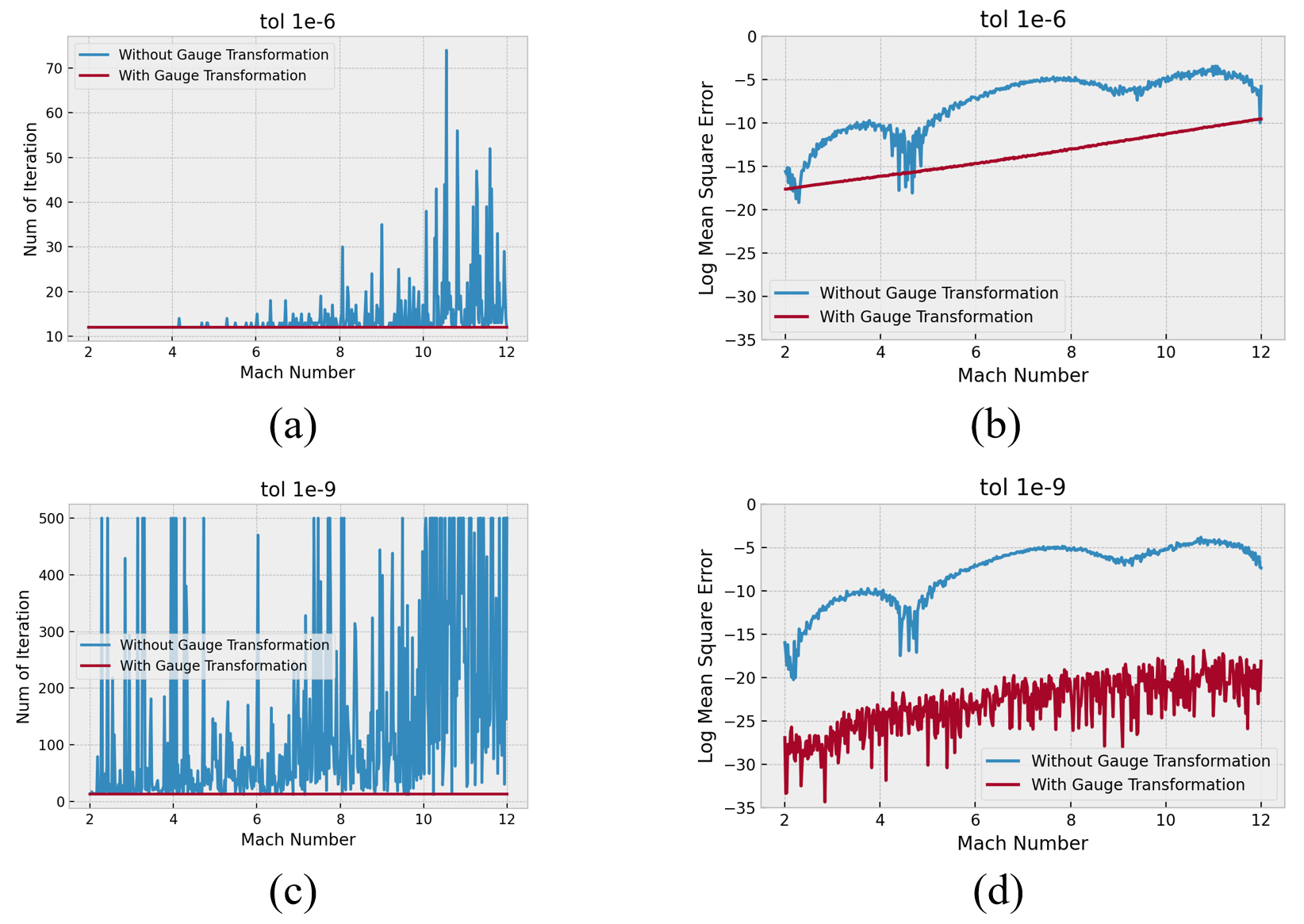}
\caption{These figures show how gauge transformation speeds up optimization at single precision. It compares the iterations needed to optimize parameters with or without gauge transformation. The goal is to find the parameters for the upper stream flow of normal shock at different Mach numbers. Figures (a) and (c) show the iterations for different Mach numbers and tolerance levels (tol=1e-6 and tol=1e-9). Without gauge transformation, optimization is slow and unstable. With Hermite gauge, optimization is fast and stable. Figures (b) and (d) show the log-mean-square error of the parameters at single precision compared to double precision. The error with Hermite gauge is much lower than without gauge transformation. These figures prove that gauge transformation makes optimization more accurate and faster.}\label{Fig3}
\end{figure}

\subsection{ Gauge transformation accelerates the convergence of optimization }

This section demonstrates the advantage of gauge transformation in accelerating optimization at single precision. Particularly, the required iterations to converge are reduced if we transform parameters and moments before optimization into the Hermite gauge. Fig.\ref{Fig3} compares the number of iterations required to optimize parameters with and without gauge transformation. These optimizations aim to determine the parameters of a series of Maxwell distributions, representing the upper flow of normal shock waves whose Mach number ranges from 2 to 12. At a specific Mach number, the density, flow velocity, and temperature of upper stream flow determine the moments $\{M_0, M_1, M_2, M_3\}$ of the sufficient statistics \eqref{Sufficient Max} of the distribution \eqref{Maximal Likelihood equation for Exponential Family Distributions: The Maxwell distribution} at $\Pr=1$ and $\sigma_{xx}=0$. The case without gauge transformation optimizes the parameters $\{\beta_0, \beta_1, \beta_2, \beta_3\}$ of the distribution \eqref{Maximal Likelihood equation for Exponential Family Distributions: The Maxwell distribution} from the known moments $\{M_0, M_1, M_2, M_3\}$ of the sufficient statistics \eqref{Sufficient Max}. The case with gauge transformation transforms both $\beta_i$ and $M_i$ into the Hermite gauge specified by the gauge parameter $\mathbf{g}_H$ as in \eqref{Hermite parameters} before optimization. The parameters in the Hermite gauge $\{\beta_0(\mathbf{g}_H), \beta_1(\mathbf{g}_H), \beta_2(\mathbf{g}_H),\beta_3(\mathbf{g}_H)\}$ of the distribution are then optimized from the known moments in the Hermite gauge $\{M_0(\mathbf{g}_H), M_1(\mathbf{g}_H), M_2(\mathbf{g}_H), M_3(\mathbf{g}_H)\}$ of the sufficient statistics \eqref{Sufficient Max}. Both cases use initialize parameter $\beta$ representing the uniform distribution with the density equals $M_0$ to ensure a fair comparison on the number of iterations. The optimization stops when it is converged at which the residual is less than the tolerance levels (tol) of residuals, or it exceeds the maximal iteration limit of 500. Any integration involved in the objective is performed over the finite domain $(u_x, u_r) \in [-30.7, 37.3] \times [0.0, 34.0]$, which is divided into $16 \times 16$ uniform blocks and integrated using 8th-order Gauss-Legendre quadrature.

\begin{figure}[!t]
\centering
\includegraphics[width=13cm]{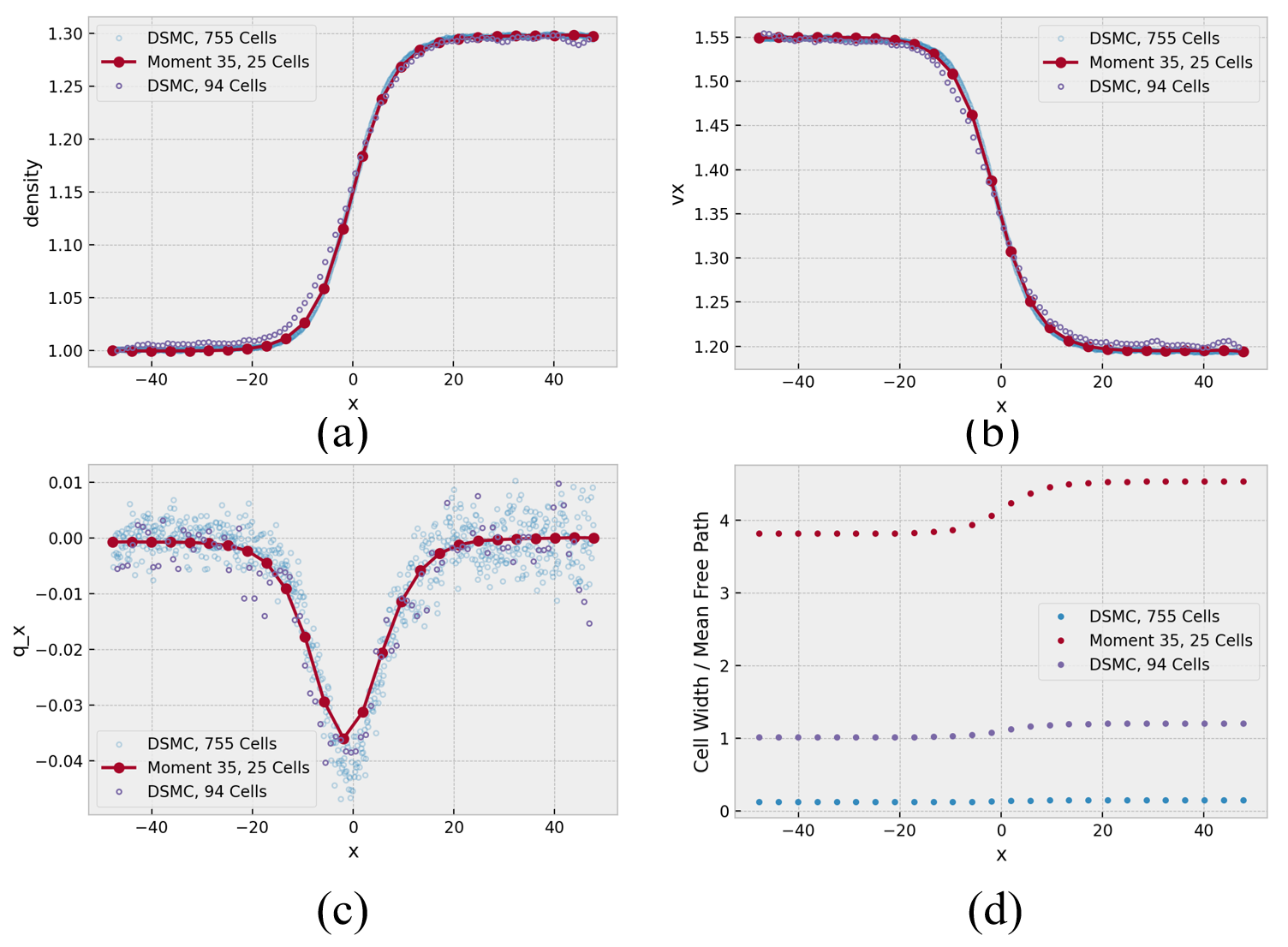}
\caption{The plots show the normal shock wave structure at Mach 1.2 computed by the MEM and DSMC methods. The MEM matches the reference DSMC (775 cells) in density, velocity, and heat flux profiles (Figures a-c), while using a coarser resolution of 4 times the mean free path. The DSMC (94 cells) deviates from the reference when the cell width is similar to the mean free path (Figure d). The MEM can capture the flow accurately with resolutions larger than the mean free path hence reduces computation costs, which is hard for mesoscopic methods like the DSMC. }\label{Fig4}
\end{figure}

Fig.\ref{Fig3}(a) and Fig.\ref{Fig3}(c) show the number of iterations required to achieve convergence for various Mach numbers and different tolerance levels (1e-6 and 1e-9) of residuals. Optimizing parameters without gauge transformation converges slowly, as indicated by the heavily fluctuating required iterations to convergence as the Mach number increases. Contrarily, optimizing parameters in the Hermite gauge converge faster with stable required iterations to convergence. Figures (b) and (d) plot the log-mean-square error of the parameters optimized at single precision w.r.t those at double precision. The log-mean-square error of parameters in the Hermite gauge is lower. It decreases further at more stringent tolerance (tol=1e-9) levels of residual while the error of parameters without gauge transformation does not. Overall, these figures demonstrated that optimization with gauge transformation is more accurate and converges faster.

\subsection{Near Continuum Shock Wave: Resolution Coarser Than the Mean Free Path is Enough} \label{continuum shock}

This section demonstrates that MEM accurately computes continuum flows with spatial resolution coarser than the mean free path of gas molecules. Fig.\ref{Fig4} studies the normal shock wave structure at Mach number 1.2, a configuration of nearly continuum flow, using MEM and the Direct Simulation Monte Carlo (DSMC) method. The MEM approach is solved with the two-step Lax-Wendroff method \eqref{The lax W}. The initial condition of the shock follows \eqref{Mx ini} with the thickness parameter $T = 15.9$. The spatial domain of simulation is $x\in[-47.7,47.7]$ divided uniformly into 25 cells. Any integration involved is performed over the finite domain $(u_x, u_r) \in [-5.4, 7.8] \times [0.0, 6.6]$, which is divided into $2 \times 2$ uniform blocks and integrated using 8th-order Gauss-Legendre quadrature. The optimization stops when it is converged at which the residual is less than $10^{-8}$, or it exceeds the maximal iteration limit of 500. 

The density, flow velocity, and heat flux profiles of the normal shock wave are depicted in Figures \ref{Fig4} (a), (b), and (c), respectively. These profiles were obtained from the MEM method and compared to those obtained from the reference DSMC computation. Figure (d) illustrates the ratio of cell width to the mean free path of gas molecules for various methods.
The results of the MEM approach are found to be in good agreement with those of the reference DSMC computation, even with a relatively low spatial resolution of 25 cells. In comparison, the results of the DSMC method (94 cells) deviate from the reference because its cell width is comparable to the mean free path of the gas molecules.

The MEM demonstrated its ability to capture the flow characteristics accurately with spatial resolutions coarser than the mean free path of the gas molecules. This ability is a significant advantage over DSMC and Discrete Velocity Methods (DVM), which require finer spatial resolution than the mean free path to simulate molecule collisions. The maximal entropy moment equation thus represents a promising tool for studying normal shock wave structures in gas dynamics.

\subsection{ The Strong Shock Wave at Mach 10: Pushing Single Precision Calculation to Its Limit}

\begin{figure}[!t]
\centering
\includegraphics[width=13cm]{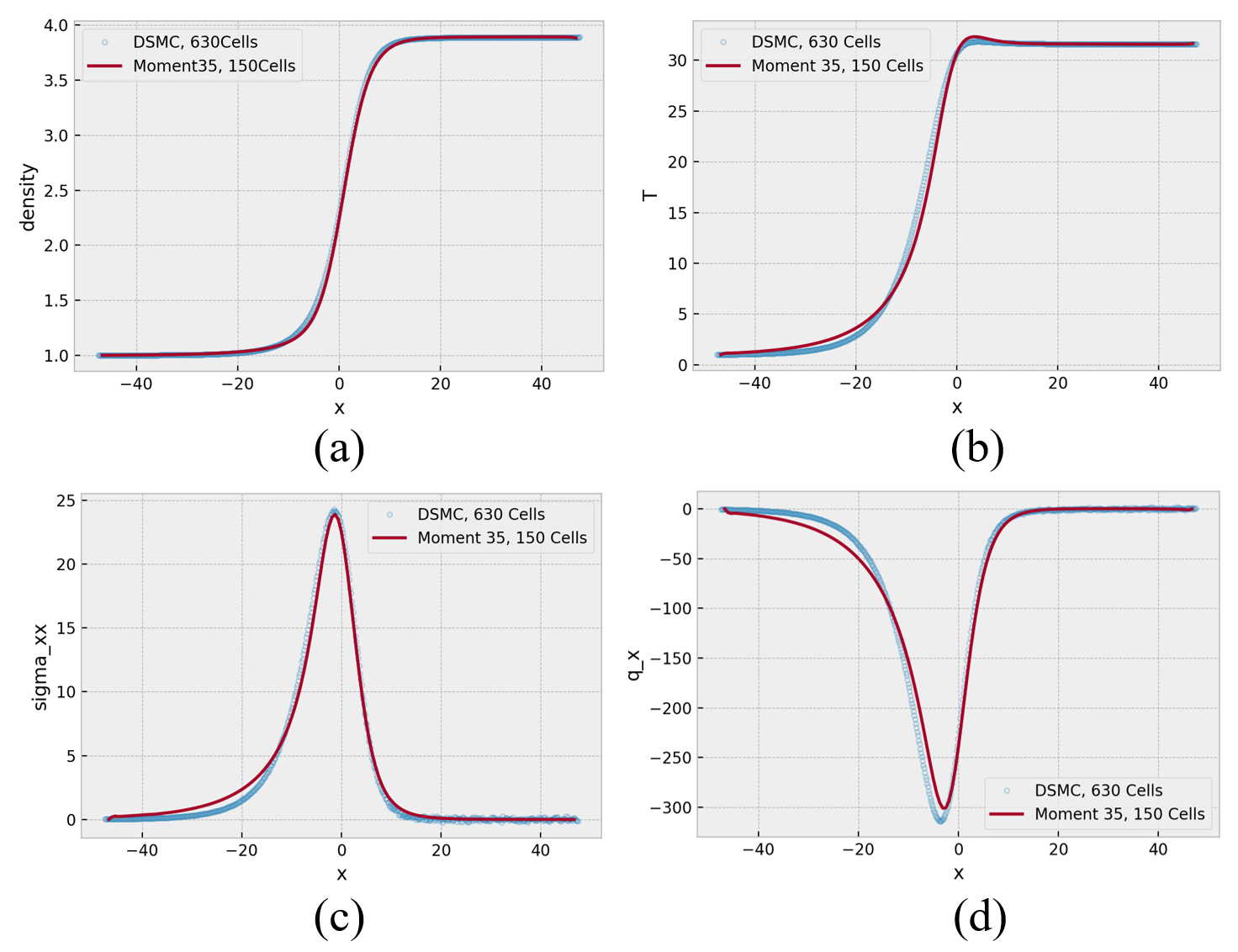}
\caption{These plots provide the first prediction of Mach 10 shock wave by MEM utilizing 35 moments of the normal shock wave at single precision, which was too ill-conditioned to be solved previously even with double precision. The density profile in (a) matches DSMC, while the temperature profile in (b) has a higher peak and amplitude near the shock. The stress and heat flux profiles in (c) and (d) also agree with DSMC at and behind the shock, but show higher amplitude in front of the shock. In addition, tiny sub-shocks appear at the upstream end of the stress and heat flux profiles as known artifacts of moment equations. These single precision results indicate that our methods greatly enhanced the numerical stability of the MEM equations, which is necessary to simulate flow at such a high Mach number.}\label{Fig5}
\end{figure}

This section shows that our methods greatly enhanced the numerical stability of the maximal entropy moment equations (MEM) by solving very strong shock waves at single precision. Fig.\ref{Fig5} presents the first prediction of the normal shock wave at Mach number 10 by MEM with 35 moments, which was too ill-conditioned to be solved previously with double precision. The MEM is solved by the FLIC scheme \eqref{The FLIC}. Fig.\ref{Fig5} (a), (b), (c), and (d) present the density, temperature, shear stress, and heat flux profiles of the normal shock wave, respectively, and are compared to the reference Direct Simulation Monte Carlo (DSMC) results. The initial condition of the shock follows \eqref{Mx ini} with the thickness parameter $T = 15.7$. The spatial domain of simulation is $x\in[-47.1,47.1]$ divided uniformly into 150 cells. Any integration involved is performed over the finite domain $(u_x, u_r) \in [-24.8, 31.4] \times [0.0, 28.1]$, which is divided into $6 \times 4$ uniform blocks and integrated using the 8th order Gauss-Legendre quadrature. The optimization stops when it is converged at which the residual is less than $10^{-8}$, or it exceeds the maximal iteration limit of 500. 

The MEM approach accurately captures the rarefied and strongly non-equilibrium gas flow at high Mach numbers. It predicts a density profile that agrees with the DSMC results. The temperature profile predicted by MEM is in approximate agreement with DSMC, with an exaggerated peak behind the shock and a slightly higher amplitude in front of the shock. Similarly, the shear stress and heat flux profiles match DSMC at and behind the shock but with higher amplitude in front of the shock. In addition, tiny sub-shocks appear at the upstream end of the stress and heat flux profiles as known artifacts of moment equations. Overall, the MEM utilizing 35 moments accurately captures the rarefied and strongly non-equilibrium gas flow at high Mach numbers. It also proved that our methods greatly enhance the numerical stability of the MEM equations for strongly non-equilibrium flow.

\subsection{Convergence and Stability} \label{Convergence and Stability}
In this section, we demonstrate the numerical convergence of our methods at single precision on Mach 4 shock wave, which is a mild-conditioned configuration enabling us to compare with previous studies \cite{Schaerer2017The3S, Schaerer2017EfficientAA}. In addition, we investigate how spatial resolution affects the stability of MEM.

We adopt the second-order FLIC scheme \eqref{The FLIC} to compute the shock profile. The initial condition of the shock follows \eqref{Mx ini} with the thickness parameter $T = 5.44$. The spatial domain of simulation is $x\in[-16.3,16.3]$ divided uniformly into cells. Any integration involved is performed over the finite domain $(u_x, u_r) \in [-22.7, 16.1] \times [0.0, 14.5]$, which is divided into $6 \times 4$ uniform blocks and integrated using the 8th order Gauss-Legendre quadrature. The optimization stops when it is converged at which the residual is less than $10^{-9}$, or it exceeds the maximal iteration limit of 500.

\begin{figure}[!t]
\centering
\includegraphics[width=13cm]{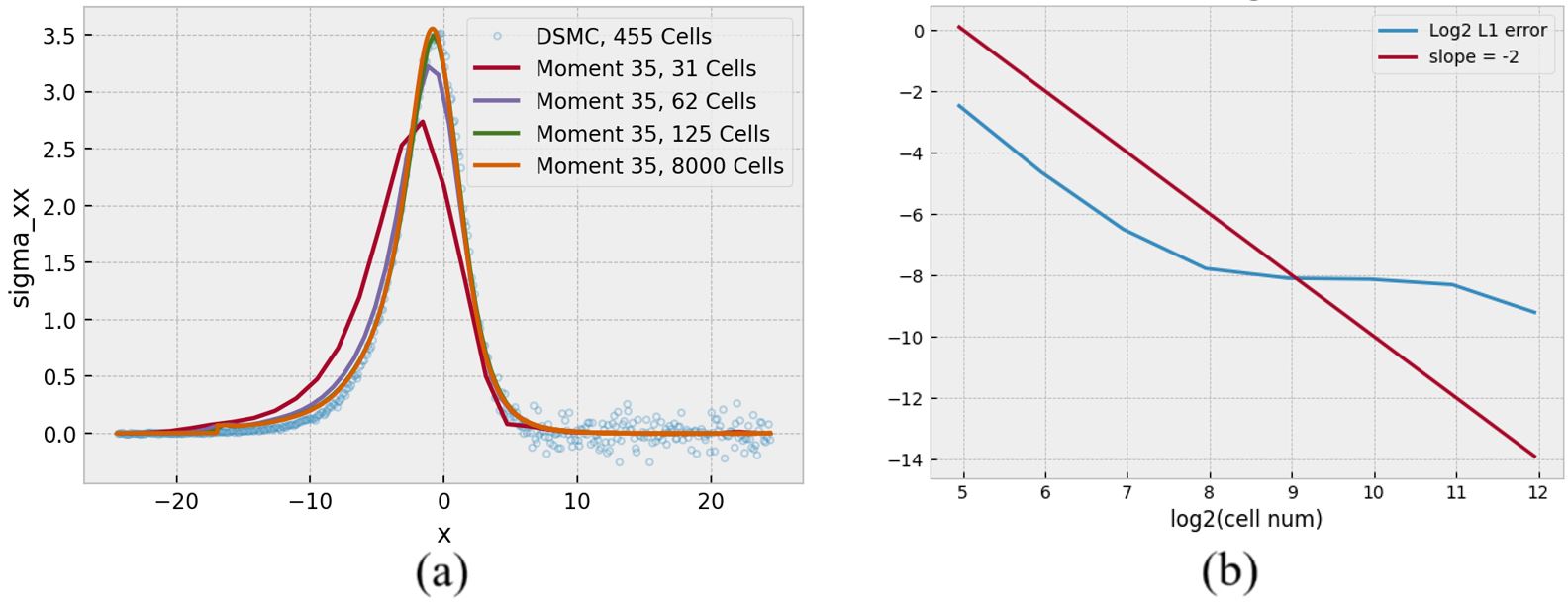}
\caption{These plots demonstrate the convergence of our algorithm. The MEM is solved by the FLIC scheme \eqref{The FLIC} at Mach number 4 at single precision, which exhibits 2nd order convergence rate before it hits the precision limit of single precision calculation at $2^8$ cells. There are sub-shocks appear at the upstream end of the stress profiles, a known artifact of moment equations. These results demonstrate that simulation with a few hundred of cells is quite satisfying with no significant deviation from results calculated with 8000 cells.}\label{Fig6}
\end{figure}

Fig.\ref{Fig6} (a) presents the convergence of the stress profile $\sigma_{xx}$ of the normal shock wave at Mach number 4, simulated by the MEM with 35 moments. It shows that our algorithm admits very fine mesh (8000 grid) similar to previous works \cite{Schaerer2017The3S, Schaerer2017EfficientAA}. It also shows that shock profiles computed with the FLIC scheme of 125 cells have no significant difference from those calculated with 8000 cells. Fig.\ref{Fig6} (b) demonstrates that our algorithm achieves the second-order convergence like the previous work \cite{Schaerer2017EfficientAA} through the L1 error of the stress profile. These results demonstrate that simulation with a few hundred of cells is quite satisfying with no significant deviation from results calculated with 8000 cells.

We did not directly compare our stress profile with the previous results presented in \cite{Schaerer2017The3S, Schaerer2017EfficientAA}. The primary reason is the difference in the collision models: while these studies utilized the vanilla BGK collision model with a Prandtl number $\mbox{Pr}=1$, the accurate Prandtl number for a monatomic gas should be $\frac{2}{3}$. In our work, we adopted the ES-BGK collision model, which accurately captures this Prandtl number for monatomic gases. Such discrepancies in the collision model can substantially alter the shock profile, rendering direct comparisons meaningless. Nevertheless, our MEM results align closely with the benchmark DSMC results, affirming the accuracy of our computations.


\begin{figure}[!t]
\centering
\includegraphics[width=13cm]{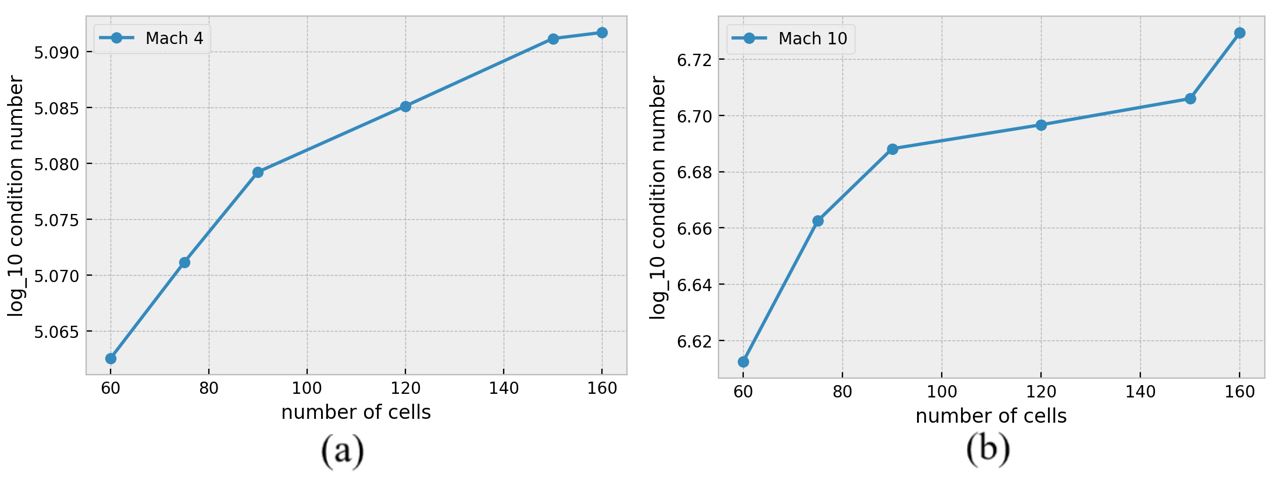}
\caption{The number of cells v.s the maximum condition number of the Jacobian \eqref{moments vs parameter jacob} encountered when simulating Mach 4 and 10 shock wave with the FLIC scheme. The condition number increases as the number of cells increases, which means the maximal entropy optimization problem \eqref{MLE optimization goal} becomes more ill-conditioned as the number of cells increases. It confirms our argument in section \ref{Resolution Limits sec} that over-refined spatial resolution damages the stability of the maximal entropy moment equations.}\label{Fig7}
\end{figure}

Fig.\ref{Fig7} plots the maximum condition numbers of the Jacobian \eqref{moments vs parameter jacob} encountered when simulating Mach 4 and 10 shocks against the number of cells. The condition number increases as the number of cells increases, which means that the maximal entropy optimization problem \eqref{MLE optimization goal} becomes more ill-conditioned as the number of cells increases. It is not a severe issue at Mach 4, whose condition number of order $10^5$ lies within the precision limit of the single precision calculation. However, it has a significant impact on the stability of the Mach 10 shock wave, whose condition number is $50$ times larger than the Mach 4 shock wave and is at the edge of the precision limit of the single-precision calculation. As a result, we observed that the simulation of the Mach 10 shock wave fails when the number of cells is greater than 160 because the Jacobian is too ill-conditioned to allow the computation of the characteristic velocity $\lambda$ in \eqref{The lax F}. This phenomenon confirms our argument in Section \ref{Resolution Limits sec} that over-refined spatial resolution damages the stability of the maximal entropy moment equations.

\begin{figure}[!t]
\centering
\includegraphics[width=13cm]{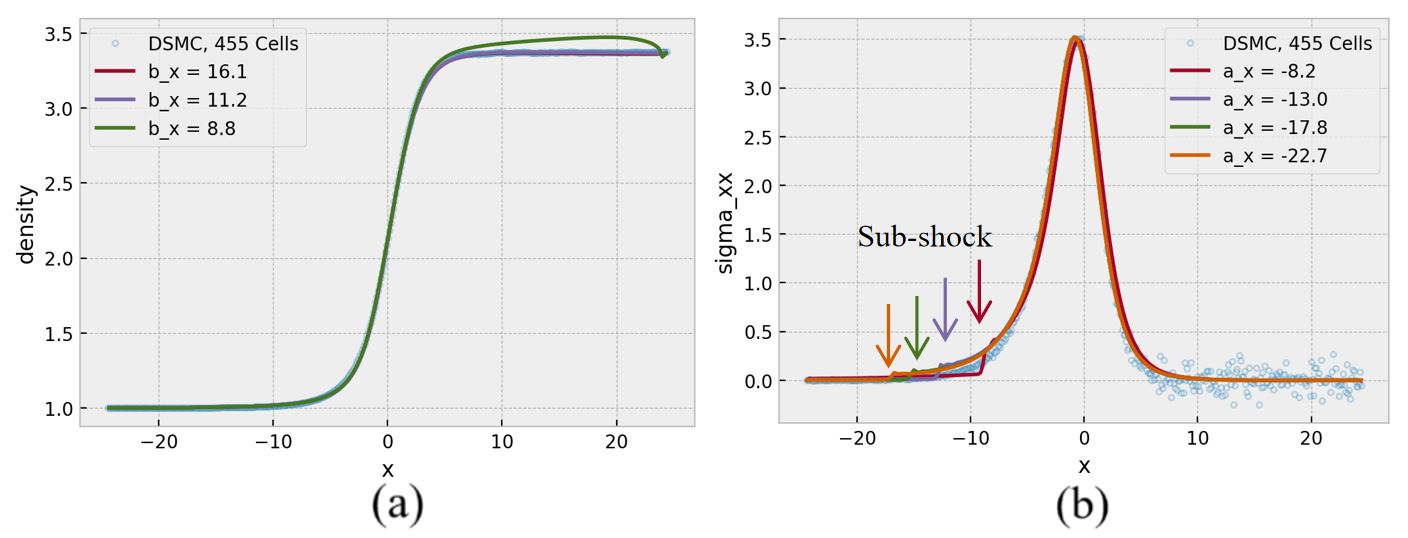}
\caption{The effect of the integration domain size $u_x\in$ $ [a_x, b_x]$ on the shock wave profile at Mach 4. Fig (a) demonstrates the change of density profile w.r.t different $b_x$. When $b_x$ is too small, irregular behavior appears in the downstream of the density profile, indicating that $b_x$ mainly effects the downstream area of the shock wave. Figure (b) shows the change of the stress profile at varies $a_x$. Discontinuity, known as the sub-shock, appears in the stress profile. The position and amplitude of the sub-shock changes with $a_x$. As $a_x$ grows, the amplitude of sub-shock decreases and the position of it moves towards the upstream direction, which is consistent with previous works. }\label{Fig8}
\end{figure}

\subsection{The domain of integration}\label{result int domain}

In this section, we explore how the shape of the shock wave at Mach 4 changes depending on the finite integration domain. We use the domain $(u_x, u_r)\in$ $ [a_x, b_x] \times [0, b_r] $ to calculate the moments of the statistics \eqref{Sufficient 35 ortho}, where $u_x$ and $u_r$ are the velocities along the x and r directions, respectively. Since shock waves are one-dimensional flows that mainly vary along the x direction, the domain boundaries $a_x$ and $b_x$ are important factors for their behavior.

To compute the shock profile, we use the 2nd-order FLIC scheme, which is given by equation \eqref{The FLIC}. The initial condition of the shock is specified by equation \eqref{Mx ini} with the parameter $T = 5.44$. The simulation domain is $x\in[-16.3,16.3]$, which is divided into 250 uniform cells. We integrate over different domains $(u_x, u_r) \in [-a_x, b_x] \times [0.0, 14.5]$. We start with $a_x = -22.7, b_x = 16.1$ and then vary $a_x$ or $b_x$ while keeping the other fixed. Each domain is divided into $6 \times 4$ uniform blocks and integrated using the eighth-order Gauss-Legendre quadrature. The optimization stops when the residual is less than $10^{-9}$ or when it reaches 500 iterations.


The density and stress profiles of the shock wave depend on the choice of the test domain $(u_x, u_r) \in [-a_x, b_x] \times [0.0, 14.5]$. The parameter $b_x$ affects the downstream region of the shock wave, while $a_x$ affects the upstream region. Fig.\ref{Fig8} (a) shows how the density profile changes with different values of $b_x$. When $b_x$ is too small, the density profile exhibits irregular behavior in the downstream region, indicating that the test domain is not large enough to capture the shock structure. Therefore, we need to choose $b_x$ sufficiently large to avoid such artifacts. Fig.\ref{Fig8} (b) shows how the stress profile changes with different values of $a_x$. The stress profile has a discontinuity, known as the sub-shock, which appears in the upstream region of the shock wave. The position and amplitude of the sub-shock vary with $a_x$. As $a_x$ decreases, the sub-shock moves closer to the upstream boundary and its amplitude decreases, which is consistent with previous works \cite{Schaerer2017The3S, Schaerer2017EfficientAA}. It is tempting to think that the sub-shock disappears as the amplitude of $a_x$ grows. However, a recent study shows that this is not the case \cite{zheng2023phase}. Therefore, to determine $a_x$, we recommend trial and error to achieve a balance between the accuracy of the shock wave and the computational efficiency.

\section{Discussion}

The paper solves the major problem preventing the practical implementation of the maximal entropy moment equations (MEM) for strongly non-equilibrium flow: the breakdown of the simulation due to numerical overflow and instability. As shown in Fig.\ref{Fig1}, we prevent numerical overflow by adopting the canonical form of the exponential family model, which introduces the partition function that can be safely computed by the log-sum-exp trick. We enhance the robustness of Newton's method in solving the parameters of the exponential family model by adopting the classical Hession modification \ref{Newton Optimization step}. This modification allows for accurate optimization in extremely ill-conditioned cases and is simple to implement, as shown in Fig.\ref{Fig2}. Moreover, we utilize gauge transformation making the optimization problem less ill-conditioned, thus accelerating the optimization process, as demonstrated in Fig.\ref{Fig3}. These methods greatly enhanced the numerical stability of the maximal entropy moment method, making it practical to simulate the strong normal shock wave at large Mach 10 ( Fig.\ref{Fig5} ) at single floating-point precision on a GPU accelerator.

MEM increases computation efficiency by coarsening the resolution in molecule velocities $\bold{u}$ of the Boltzmann equation. It compresses the three dimensions of molecule velocities into a few moments, squeezing the 6-dimension Boltzmann equation into 3-dimension moment equations that are more affordable. This reduction in dimension distinguishes MEM from DSMC and DVM, in which molecule velocities are necessary dimensions. Although this paper uses numerical integration over the molecule velocities as a makeshift estimator of moments and fluxes, it does not make molecule velocities necessary in MEM as those in the DSMC and DVM methods. On the one hand, the MEM requires much fewer sample points in molecule velocities than DVM or DSMC. It is because the exponential family distribution, characterized by a few parameters, is much smoother than the one-particle distribution computed by DVM or DSMC and hence requires fewer sample points to integrate. On the other hand, it is possible to completely eliminate numerical integration when evaluating moments and fluxes because the relation among moments, parameters, and fluxes of exponential family distribution is deterministic hence is suitable to be approximated by machine learning techniques. Overall, the maximal entropy moment equations increase the computational efficiency by reducing the 6-dimension Boltzmann equation into 3-dimension moment equations.

Another advantage of MEM over DSMC and DVM is it accurately captures near-continuum flow with spatial resolution several times coarser than the mean free path of gas molecules. It is because MEM relies only on the invariant manifold assumption described in section \eqref{Maximal Entropy Moment Equations} instead of simulating gas molecule collisions like DSMC and DVM, which require finer spatial and temporal resolution than the mean free path and time of molecules. Specifically, DSMC randomly collides particles within the same cell. This is erroneous for cells wider than the mean free path, where particles with gaps larger than the mean free path seldom collide. DVM uses the Boltzmann equation’s collision term, which assumes less than one collision per molecule pair. This is invalid for cells wider than the mean free path, across which molecules undergo multiple collisions. Unlike DSMC and DVM, bigger cells make MEM's invariant manifold assumption more accurate because they reduce the deviation of molecule velocity distribution from equilibrium. In summary, MEM has a clear advantage over DSMC and DVM on near-continuum flow because the invariant manifold assumption is valid for spatial resolution coarser than the mean free path.

However, MEM is less effective than DSMC and DVM for highly rarefied gas flows. Such flow is strongly non-equilibrium and needs many high-order polynomial moments to model. High-order polynomial moments generally lead to high characteristic speed, which significantly shrinks the time step size of numerical solvers. Consequently, MEM requires more time steps than DSMC or DVM to simulate highly rarefied gas flows. Overall, MEM is efficient for continuum flow and costly for rarefied flow, unlike DSMC and DVM, which are the opposite.

Another difficulty MEM encountered is its resolution limits due to the invariant manifold assumption, as discussed in section \ref{Resolution Limits sec}. These resolution limits set the minimum time step and spatial mesh spacing MEM can adapt, beyond which MEM no longer approximates the Boltzmann equation. In practice, the existence of such limits is evidenced by numerical instability or the breakdown of the simulation using an over-refined spatial mesh. The direct cause of instability or breakdown is an explosion in the maximal characteristic speed. The underlying reason is that, as the cell width decreases, the average velocity distribution over a cell deviates more from equilibrium, making the Jacobian between fluxes and moments more ill-conditioned. Such ill-conditioned Jacobian could be mitigated by shrinking the velocity domain of numerical integration at the cost of an enlarged sub-shock. Respecting the resolution limit by using spatial cells wider than the mean free path serves as a workaround. However a general and complete solution may require dissipative regularization terms such as viscosity into MEM, but their specific formulation needs more investigation.

In conclusion, this paper overcomes the main challenge for using MEM to strongly non-equilibrium flow: the simulation failure due to numerical overflow and instability. It achieves this by applying the canonical form of the exponential family model, a modified Newton’s optimization, and gauge transformations. We succeeded in simulating the strong normal shock wave at Mach 10 with MEM and 35 moments for the first time, using single floating-point precision on a modern GPU. We also point out that overrefined spatial mesh is unnecessary and undermines stability. This paper makes the maximal entropy moment method practical for simulating very strong normal shock waves on modern GPUs at single floating-point precision, with significant stability improvement compared to previous methods.

\section*{Acknowledgments}
We thank Dr. Yuan Lan (Hong Kong University of Science and Technology) for her comments on the manuscript.

\appendix
\section{Collision Model of the Boltzmann Equation}\label{sec AB}
In addition to density, flow velocity and temperature, we could define the pressure tensor $P$ and the heat flux vector $\mathbf{q}$
\begin{equation} \label{Pressure, Stress, and Heat flux}
\begin{split}
    P_{\alpha\beta} &= p \delta_{\alpha\beta} + \sigma_{\alpha\beta} = \int m c_\alpha c_\beta f(\bold{u}) d^3 \bold{u}\\
q_\alpha &= \int \frac{m}{2} c_\alpha \bold{c}^2 f(\bold{u}) d^3 \bold{u};\quad \alpha,\beta = 1,2,3,
\end{split}
\end{equation}
in which $c_\alpha$ be the $\alpha$th component of peculiar velocity $\bold{c}$, $p$ is the pressure, and $\sigma_{\alpha\beta}$ is the shear stress tensor which is traceless.

The collision term $Q(f,f)$ of the Boltzmann equation is complex and computationally expensive. Instead, we adopt simplified collision models such as the well-known BGK collision model \cite{Bhatnagar1954AMF}, replacing the collision term $Q(f,f)$ with a linear functional $C(f)$
\begin{equation} \label{The BGK model}
C(f) = \frac{\Tilde{f}-f}{\tau},
\end{equation}
in which $\tau$ is the relaxation time, $\Tilde{f} = f_0$ is the Maxwell distribution with the same density, flow velocity, and temperature with $f$. The BGK model has the Prandlt number fixed at $\Pr=1$. A more powerful model of $\Tilde{f}$ with an adjustable Prandlt number is the elliptic statistical model \cite[p.97]{TheBoltzmannEquationAndItsApplications}
\begin{equation} \label{The ES BGK model}
\begin{split}
\Tilde{f} &= \frac{n }{ \sqrt{\det 2\pi  \boldsymbol{\Sigma}} } \exp\left[ -\frac{1}{2} (\mathbf{u}-\mathbf{v})^T\boldsymbol{\Sigma}^{-1}(\mathbf{u}-\mathbf{v}) \right] \\
\boldsymbol{\Sigma} &= \{\Sigma_{\alpha \beta}\} =\frac{1}{\Pr}\frac{ k_B T}{m }\delta_{\alpha \beta} + (1- \frac{1}{\Pr})\frac{P_{\alpha \beta}}{nm};\quad \Pr > \frac{1}{2}
\end{split}
\end{equation}	
in which $\delta_{\alpha \beta}$ represents the identity matrix of the size $3\times 3$, $n$, $\mathbf{v}$, $T$ and $P_{\alpha \beta}$ are the number density, flow velocity, temperature, and pressure tensor of the one-particle distribution $f$. 

This paper considers the monatomic gas of Maxwell molecules with the Prandlt number $\Pr = \frac{2}{3}$. Maxwell molecules are characterized by their polynomial interacting potential $U(r) = k r^{-4}/4 $ between two molecules separated at a distance $r$. The viscosity coefficient $\mu$ of Maxwell molecule gas is proportional to the temperature of the gas \cite[p.68]{MolecularGasDynamicsAndTheDirectSimulationOfGasFlows}
\begin{equation} \label{Maxwell viscosity coef}
	\mu = \sqrt{\frac{m}{2k}}\frac{ 2 k_B T}{ 3A_2 }, \quad A_2 = 1.37035
\end{equation}
It determines the relaxation time $\tau$ of the elliptic statistical BGK model by the following relation
\begin{equation} \label{The ES BGK viscosity}
\frac{\mu}{\Pr} = n k_B T \tau
\end{equation}	

\section{Exponential Family Model of the 35 Moment System} \label{sec AM35}

This paper adopts the 35-moment system \cite{Schaerer2017EfficientAA}, whose sufficient statistics for one-dimensional flow along the $x$ direction are as follows
\begin{equation}
\label{Sufficient 35 dim}
\begin{split}
\{\phi_i(\bold{u}), &i=0, \cdots, 8\}
    =\{1, {u}_x,{u}_x^2, {u}_r^2, 
    {u}_x^3, {u}_x^4, {u}_r^4, {u}_x {u}_r^2 ,{u}_x^2{u}_r^2 \}
\end{split}
\end{equation}
in which $u_r = \sqrt{u_y^2 + u_z^2}$. This moment system includes all monomials of $u_x$ up to the 4th order, which is capable of modeling both stress (2nd order) and heat flux (3rd order) in the $x$ direction. Besides, the 4th order monomials $u_x^2, u_r^2$ in the moment system are essential to ensure that the density distribution vanishes at infinity. 

Under proper non-dimensionalization \eqref{Non-dim}, the 35-moment system could be rewritten into the following orthogonality form: 
\begin{equation}
\label{Sufficient 35}
\begin{split}
\{\phi_i(\bold{u}), &i=0, \cdots, 8\}
    =\{1, {u}_x,\frac{ {u}_x^2-1}{\sqrt{2}},\frac{ {u}_r^2}{2} -1, \\
    &\frac{ {u}_x^3-3 {u}_x}{\sqrt{6}},\frac{ {u}_x^4-6 {u}_x^2+3}{2 \sqrt{6}},\frac{1}{8}  {u}_r^4- {u}_r^2+1,\frac{1}{2}  {u}_x ( {u}_r^2-1),\frac{(  {u}_x^2 -1)(  {u}_r^2-2)}{2 \sqrt{2}}\}
\end{split}
\end{equation}

 In addition, we are also interested in sufficient statistics $\psi_i(\bold{u})$ as a subset of \eqref{Sufficient 35}
\begin{equation}
\label{Sufficient Max}
\{\psi_i(\bold{u}), i=0, \cdots,3\} = \{1, u_x, u_x^2, u_r^2\},
\end{equation}
which describes the elliptic statistical model \eqref{The ES BGK model}. Notably, the elliptic statistical model \eqref{The ES BGK model} for one-dimensional flow with flow velocity $\bold{v} = \{v_x,0,0\}$ could be described by statistics \eqref{Sufficient Max} in the canonical form as
\begin{equation} \label{Maximal Likelihood equation for Exponential Family Distributions: The Maxwell distribution}
	\begin{split}
  \Tilde{f}(t,\bold{x},\bold{u})&=n \exp \left( \frac{ \Pr  v_x}{A}\psi_1(\mathbf{u})-\frac{ \Pr  }{2 A}\psi_2(\mathbf{u})-\frac{ \Pr  }{B}\psi_3(\mathbf{u}) +\log \left( \frac{ \Pr^3 }{ 2 A B^2\pi^3 }\right)^\frac{1}{2}-\frac{ \Pr  v_x^2}{2A} \right)\\
  A &=\frac{k_B n \Pr  T+(\Pr -1) \sigma_{xx} }{nm};\quad  B = \frac{2 k_B n \Pr  T+ (1-\Pr) \sigma_{xx}}{nm},
	\end{split} 
\end{equation}
in which $m$ is the molecule mass, $k_B$ is the Boltzmann constant, $n$, $v_x$, $T$, $\sigma_{xx}$ are the number density, flow velocity, temperature, and the first component of the shear stress tensor $\sigma$ with their time and spatial dependency omitted. 

\section{Properties of Exponential Family Distribution} \label{sec A opt}

\textbf{Moment matching} The distribution $f_{\boldsymbol{\beta}}$ having the matching moment with $f$ refers to that they have the same moment of sufficient statistics:
\begin{equation} \label{Maximal Likelihood equation for Exponential Family Distributions: exp family moments matching}
   \int  \phi_i(\bold{u})f(t, \bold{x}, \bold{u}) d^3\mathbf{u} = \int \phi_i(\bold{u}) f_{\boldsymbol{\beta}}(\mathbf{u};n(t, \bold{x}), \boldsymbol{\beta}(t, \bold{x})) d^3\mathbf{u};\quad i=0,\cdots,M.
\end{equation} 

\textbf{Minimal Distance} The distribution $f_{\boldsymbol{\beta}}$ specified by moment matching \eqref{Maximal Likelihood equation for Exponential Family Distributions: exp family moments matching} is "closest" because it is the optimal solution of minimizing the Kullback–Leibler divergence $\mbox{KL}( f \| f_{\boldsymbol{\beta}} )$ between $f$ and $f_{\boldsymbol{\beta}}$ \begin{equation} \label{Maximal Likelihood equation for Exponential Family Distributions: mle exp family}
	\begin{split}
\min_{\boldsymbol{\beta}} \mbox{KL}( f \| f_{\boldsymbol{\beta}} ) = \int &f( t,\bold{x}, \mathbf{u} )\log \frac{f( t,\bold{x}, \mathbf{u} )}{f_{\boldsymbol{\beta}}(\mathbf{u}; \boldsymbol{\beta}(t, \bold{x}))}  d^3\mathbf{u} \\
		s.t\ \ \int  f( t,\bold{x}, \mathbf{u} ) d^3\mathbf{u}&=\int  f_{\boldsymbol{\beta}}(\mathbf{u};\boldsymbol{\beta}(t, \bold{x})) d^3\mathbf{u}.
	\end{split}
\end{equation}Moreover, $f_{\boldsymbol{\beta}}$ is the maximal likelihood estimation of the one-particle distribution $f$ because minimizing the Kullback–Leibler divergence is equivalent to maximizing the likelihood \cite{Wasserman2004MLE}. 

\textbf{Maximal Entropy} The entropy of $f_{\boldsymbol{\beta}}$ is always larger than the entropy of $f$ because $f_{\boldsymbol{\beta}}$ have the maximal entropy given the moments of sufficient statistics. It is because $f_{\boldsymbol{\beta}}$ is the optimal solution to the maximal entropy problem \cite{MAL-001ExpFamilyMaxent}
\begin{equation} \label{Maximal Likelihood equation for Exponential Family Distributions: max ent exp family}
	\begin{split}
		\max_{g} \int &-g( t,\bold{x}, \mathbf{u} )\log g( t,\bold{x}, \mathbf{u} ) d^3\mathbf{u} \\
		s.t\ \ \int \phi_i(\bold{u}) f(t, \bold{x}, \bold{u}) d^3\mathbf{u} &= \int  \phi_i(\bold{u})g( t,\bold{x}, \mathbf{u} ) d^3\mathbf{u} ; i=0,\cdots,M
	\end{split}
\end{equation} hence possesses the maximized entropy among distributions having the same moments of sufficient statistics. In conclusion, for one-particle distribution $f$ not in the exponential family, we can find its "closest" distribution $f_{\boldsymbol{\beta}}$ in the exponential family with the same moments of sufficient statistics but higher entropy via moment matching \eqref{Maximal Likelihood equation for Exponential Family Distributions: exp family moments matching}.

\textbf{Solving Parameters $\boldsymbol{\beta}$ } In practice, we find $f_{\boldsymbol{\beta}}$ 
 by solving its parameters $\boldsymbol{\beta}$ from a variant of the optimization problem \eqref{Maximal Likelihood equation for Exponential Family Distributions: mle exp family} as the following
\begin{equation}
\label{MLE optimization goal}
	\begin{split}
\min_{\boldsymbol{\beta}} &\quad L( \boldsymbol{\beta}; \bold{M}, \boldsymbol{\phi}) = \log Z(\boldsymbol{\beta}; \boldsymbol{\phi}) - \beta_i \frac{M_i}{n} - n^2\left(\log \beta_0 -\frac{\beta_0}{n} \right);\quad i=1,\cdots, M,
	\end{split}
\end{equation}                                                                                                                  
in which $\{M_i, i=0,\cdots,M\}$ are moments of $f$ defined in \eqref{Maximal Likelihood equation for Exponential Family Distributions: exp family moments}, $n = M_0$ is the number density, $Z(\boldsymbol{\beta}; \boldsymbol{\phi})$ is the partition function defined in \eqref{Maximal Likelihood equation for Exponential Family Distributions: exp family}. The optimization problem \eqref{MLE optimization goal} is derived from the maximal likelihood problem \eqref{Maximal Likelihood equation for Exponential Family Distributions: mle exp family} by substituting $f_{\boldsymbol{\beta}}$ with its canonical form \eqref{Maximal Likelihood equation for Exponential Family Distributions: exp family}. It differs from but is equivalent to the optimization objectives \cite[Eq. 9]{Alldredge2013AdaptiveCO}\cite[Eq. 22]{Schaerer2017EfficientAA} in previous works, with the difference caused by adopting the canonical form \eqref{Maximal Likelihood equation for Exponential Family Distributions: exp family} of the exponential family model. 

The Hessian of the optimization problem \eqref{MLE optimization goal} is 
\begin{equation}
\label{MLE optimization goal hessian}
H_{ij} = \frac{\partial^2 L}{\partial \beta_i \partial \beta_j} = \begin{cases}
n^2/\beta_0^2,\quad &i = j = 0 \\
\left< \phi_i(\bold{u}) \phi_j(\bold{u}) \right>,\quad &i\ge1, j \ge 1\\
0,\quad & \mbox{otherwise} \\
\end{cases},
\end{equation}
in which $\left< \phi_i(\bold{u}) \phi_j(\bold{u}) \right>$ is the covariance between sufficient $\phi_i(\bold{u})$ and $\phi_j(\bold{u})$ defined as
\begin{equation}\label{correlation}
\begin{split}
    \left< \phi_i(\bold{u}) \phi_j(\bold{u}) \right> &= \frac{1}{\beta_0} \int \phi_i(\bold{u}) \phi_j(\bold{u})f_{\boldsymbol{\beta}}(\mathbf{u};n, \boldsymbol{\beta})d^3\mathbf{u} \\&- \frac{1}{\beta_0^2} \int \phi_i(\bold{u})f_{\boldsymbol{\beta}}(\mathbf{u};n, \boldsymbol{\beta})d^3\mathbf{u} \int \phi_j(\bold{u}')f_{\boldsymbol{\beta}}(\mathbf{u}';n, \boldsymbol{\beta})d^3\mathbf{u}'.
\end{split}
\end{equation}
This Hessian is crucial in analyzing and solving the optimization problem \eqref{MLE optimization goal}.

\section{Simulating the Normal Shock Wave} \label{shock wave sims}

The shock wave thickness is a classical benchmark for rarefied gas simulation \cite{Bird1994}[Sec 12.11]. A shock wave in gas is a disturbance that propagates faster than the speed of sound. It is characterized by sudden changes in the density, velocity, and temperature of the gas. Such changes happen across surfaces known as the shock front, whose thickness has the same order as the mean free path of gas molecules. The mean free path of gas molecules is small in dense gas, in which we could neglect the thickness of the shock front. Contrarily, we shall not disregard the thickness of the shock front in rarefied gas. Strong non-equilibrium dynamics happen within the shock front, making shock waves a nice benchmark to examine the hydrodynamic model for rarefied gas dynamics.

Shock waves can be normal, oblique, and bow shocks, according to the geometry of the flow. We consider normal shocks, in which the shock front is perpendicular to the gas flow velocity. Normal shocks are axial symmetric w.r.t the flow velocity, thus are one-dimensional flows. In addition, Normal shocks are stationary in the inertial frame moving with the shock front. These properties significantly reduce the complexity of normal shocks.

The shock front splits a shock flow field into the upstream and downstream regions. Without loss of generality, we consider uniform upstream and downstream flows with density $\rho_{\pm}$, velocity $v_{\pm}$, and temperature $T_{\pm}$, in which $-$ represent upstream and vice versa. However, these quantities are not independent due to the conservation of mass, momentum, and energy. There is only one degree of freedom among these quantities despite the change of units of measure, characterized by the upstream Mach number
\begin{equation}
     M_{-} = v_{-}\sqrt{\frac{m}{\gamma k_B T_{-}}},
\end{equation}
in which $m$ is the mass of gas molecules, $\gamma=\frac{5}{3}$ is the heat capacity ratio, $k_B$ is the Boltzmann constant. The upstream Mach number uniquely determines the ratio between upstream and downstream density, velocity, and temperature, known as the Rankine-Hugoniot condition \cite{Landau1987} 
\begin{equation} \label{RH condition}
\frac{\rho_+}{\rho_-}=\frac{v_-}{v_+} = \frac{(\gamma+1)M_-^2}{(\gamma-1)M_-^2+2} \quad 	\frac{T_+}{T_-}=\frac{\left[2\gamma M_-^2-(\gamma-1)\right]\left[(\gamma-1)M_-^2+2\right]}{(\gamma+1)^2M_-^2}.
\end{equation}
In addition to the upstream Mach number, the upstream Knudsen number
\begin{equation}
     \mbox{Kn}_- = \sqrt{\frac{ \pi m}{2  k_B T_-}}\frac{\mu_-}{\rho_- \Delta x},
\end{equation}
in which $\mu_-$ is the upstream viscosity coefficient and $\Delta x$ is the unit length, determines the thickness of the shock wave. Normal shock waves with the upstream Mach number, Knudsen number, and unit length specified could be simulated with arbitrary non-dimensionalization of density, velocity, and temperatures. This paper adopts the following non-dimensionalized quantities
\begin{equation}\label{Non-dim}
m = 1\quad \rho_-=1 \quad v_- = \sqrt{\gamma} M_- \quad T_-=1 \quad \Delta x= 1 \quad \mbox{Kn} = 1.
\end{equation}
These quantities completely specify the parameters of the normal shock waves.

In practice, we simulate normal shock waves on a finite spatial domain $[x_-,x_+]$. The boundary condition $\mathbf{M}_-, \mathbf{M}_+$ at the positions $x_-,x_+$ for the moment equations \eqref{The ME equation} are moments of the upstream and downstream Maxwell distributions with corresponding densities, flow velocities, and temperatures. Specifically, the boundary condition of upstream and downstream moments $\mathbf{M}_-, \mathbf{M}_+$ are moments of the statistics \eqref{Sufficient 35 ortho} in the corresponding gauge. In addition, the initial condition $\mathbf{M}(x)$ is the linear interpolation of $\mathbf{M}_-, \mathbf{M}_+$ of the form (Modified from $2T$ to $T/2$)
\begin{equation}\label{Mx ini}
\mathbf{M}(x) = \left(\mathbf{M}_+ - \mathbf{M}_-\right)\left(  \frac{ \tanh\left(\frac{x}{T/2}\right) - \tanh\left(\frac{x_-}{T/2}\right) }{\tanh\left(\frac{x_+}{T/2}\right) - \tanh\left(\frac{x_-}{T/2}\right)} -\frac{1}{2}\right) + \frac{\mathbf{M}_+ + \mathbf{M}_- }{2}
\end{equation}
in which $T$ is the thickness parameter.

Besides, in our simulation of shock waves, we compute moments of the statistics \eqref{Sufficient 35 ortho} by integrating $(u_x, u_r)$ in a finite integration domain of the form $(u_x, u_r)\in$ $ [a_x, b_x] \times [0, b_r] $. This integration domain is determined using the following formula:
\begin{equation}
\begin{split}
a_x & = \min \left( v_+ - n_a \sqrt{\frac{k_B T_+}{m}}, v_- - n_a\sqrt{\frac{k_B T_-}{m}} \right), \\
b_x & = \max \left( v_- + n_b \sqrt{\frac{k_B T_-}{m}}, v_+ + n_b \sqrt{\frac{k_B  T_+}{m}} \right), \\
b_r & = \max \left( n_r  \sqrt{\frac{k_B  T_-}{m}}, n_r  \sqrt{\frac{k_B T_+}{m}} \right).
\end{split}
\end{equation}
in which $n_a$, $n_b$, and $n_r$ are set with integers according to the specific scenario. It turns out that the value of $n_a$ significantly affects the position of sub-shocks as shown in Section \ref{result int domain}. Therefore, we recommend setting $n_b = n_r \ge 4$ and $n_a > n_b$ to simulate shock waves.

\section{The DSMC Calculation of Shock Wave Structure}\label{secA1}

We use the DSMC1S program by A.Bird \cite{Bird1994MolecularGDF} to simulate the structure of 1D shock wave for Maxwell molecule gas with molecule mass $m=6.64\times 10^{-26} kg$ and specific heat ratio $\gamma = \frac{5}{3}$. 

The geometry of the computation domain is one-dimensional gas of a unit cross-section divided by uniform computation cells. The width $\Delta x$ and the number of computation cells $N_c$ varies for different upstream flow Mach numbers. Particularly, we set the total spatial span $N_c \Delta x$ of the computation domain to be $6$ times the shock wave thickness. We also set the widths $\Delta x$ of each computation cell to be $0.15$ times of the minimum mean free path between upstream and downstream mean free paths.

The DSMC1S program uses discontinuous initial conditions. For computation cells upstream of the origin, the DSMC simulation particles are drawn from the equilibrium distribution corresponding to the upstream density, velocity, and temperature. Particularly, we set the upstream number density to $n_1 = 10^{20} m^{-3}$ and the upstream temperature $T_1$ to $293K$. The upstream flow velocity $v_1$ are set w.r.t to the upstream Mach number $M_1$, which is defined as.
\begin{equation}
    M_1 = \frac{v_1}{\sqrt{\frac{\gamma k_B T_1}{m}}}
\end{equation}
For cells downstream of the origin, particles are drawn from equilibrium corresponding to the downstream number density $n_2$, velocity $v_2$, and temperature $T_2$. These quantities are computed by the Rankine-Hugoniot relation
\begin{equation} \label{Shock Waves in a polytropic gas: ratio in mach number}
	\begin{split}	\frac{n_2}{n_1}&=\frac{v_1}{v_2} = \frac{(\gamma+1)M_1^2}{(\gamma-1)M_1^2+2}\\
	\frac{T_2}{T_1}&=\frac{\left[2\gamma M_1^2-(\gamma-1)\right]\left[(\gamma-1)M_1^2+2\right]}{(\gamma+1)^2M_1^2}.
	\end{split}
\end{equation}
Both the upstream and downstream parts have the same size in the computational domain.  Consequently, the shock center is expected to be the origin. 

The boundary condition for DSMC1S.FOR differs for up/downstream flows. The upstream boundary condition processes fixed the density, velocity, and temperature and are at equilibrium. At each time step, particles enter the computational domain at upstream according to a fixed flux rate computed for equilibrium distribution \cite[Eq 4.22]{Bird1994}, which molecules outside the boundary are removed. However, the downstream boundary condition utilizes the "moving piston" boundary condition to reduce the fluctuations in the number of molecules in the computational domain. The "moving piston" is a specularly reflecting solid surface that moves at downstream flow speed. It affects little on the mean flow property since it matches the downstream flow speed. Moreover, fluctuations in downstream flow could not propagate upstream hence no not affecting upstream flow. The specularly reflecting surface reduces the variance in the number of molecules leaving the computational domain by fixing the mean molecules' velocity leaving the computational domain to the surface velocity. The `moving piston' differ from a normal specularly reflecting solid surface since it "jumps" back to its original position after each time step. The reason is that a normal reflecting surface moving at downstream flow speed will leave the computational domain since it is not stationary. As a solution, after each time step, we require the surface to jump back to its original position at the boundary of the computational domain. Consequently, molecules outside the surface after the jump backstage are also removed.

These boundary conditions are not enough to ensure the shock center is located at the origin. It is because the shock center location fluctuates with the total number of molecules inside the computational domain. The upstream boundary condition has no control over the molecules leaving the computational domain. The downstream boundary condition also introduces fluctuations at the "jump back" stage. As a solution, an additional function $STABIL$ appears in DSMC1S.FOR to make sure the shock wave is centered at the origin. The $STABIL$ achieves this by ensuring the total amount of the particle in the computation region lies in a small range. If the total amount of particles deviates above a threshold, then all molecules are displaced by the same distance. The molecules move outside the boundary are removed, while those originally within the displacement distance from the boundary are duplicated. Since the upstream/downstream density differs from each other, such displacement increase/decrease the total number of molecules in the computational domain according to the sign of the displacement. By manipulating the displacement distance, the $STABIL$ controls the total number of molecules in the computational domain hence the shock wave center location.

The total number of real molecules $N$ in the computation domain is computed by averaging the upstream and downstream number density since the shock wave centered is at the origin.
 \begin{equation} \label{DSMC shock wave: Total molecule number}
 	N = \frac{n_1}{2} \left(1 + \frac{(\gamma+1)M_1^2}{(\gamma-1)M_1^2+2} \right)  N_c \Delta x
 \end{equation}
The total number of DSMC simulation particles $N_p$ in the computation domain is determined by 
 \begin{equation} \label{DSMC shock wave: Total sim particle number}
 	N_p = \frac{N}{N_{eff}},
 \end{equation}
in which $N$ is the total number of real molecules and $N_{eff}$ is the effective number of real molecules each simulation particle represents. In our computation, $N_{eff}$ is chosen to ensure the number of simulation particles in upstream computation cells $\frac{n_1\Delta x}{N_{eff}} = 50$.



The molecular model is crutial in DSMC calculations. It describes how two molecule collide with each other and determines the viscosity of the gas. The molecular model gives the relation between two characteristic quantities of classical binary collision problem \cite{Bird1994Molecularcollide, griffiths_schroeter_2018collide,landau1982mechanics}: the impact parameter $b$ and scattering angle $\beta$. One of the typical molecular model used in DSMC is the variable soft sphere model \cite{Bird1994MolecularMModel}

\begin{equation} \label{VHS model: The deflection angle}
	\beta = 2 \cos^{-1}((\frac{b}{d})^{\frac{1}{\alpha}})
\end{equation}
in which $d$ is the effective diameter of the gas molecules and $\alpha$ is a parameter mainly effecting the diffusion coefficient. Therefore we use the value $\alpha=2.13986$ in our DSMC computation to make the ratio between viscosity and momentum transfer cross section \cite[Eq 2.28, 2.29]{Bird1994CrossSections} matching the result for Maxwell molecules. The effective diameter $d$ varies with the relative velocity between colliding molecules according to \cite[Eq 4.63]{Bird1994MolecularEffd}
\begin{equation} \label{DSMC: the effective diameter, calculation, final}
	d = d_{ref}(\frac{(2k_B T_{ref}/(\frac{1}{2}m v_r^2))^{\omega-1/2}}{\Gamma(5/2-\omega)})^{1/2}
\end{equation}
in which $m$ is the mass of gas molecule, $v_r$ is the relative velocity between the two colliding molecules, $\Gamma$ represents the Gamma function, $T_{ref}$ is the reference temperature, $d_{ref}$ is the reference molecule diameter, and $\omega$ is a parameter determines how viscosity coefficient changes w.r.t temperature. In our computation we use the default value $T_{ref}=293K$, and $d_{ref}=4.092 \times 10^{-10}m$ in the DSMC1S program. Note that \eqref{DSMC: the effective diameter, calculation, final} differ from the equation in \cite{Bird1994MolecularEffd} since they use the reduced mass $m_r=\frac{1}{2}m$ instead of molecule mass $m$ in our case. The parameter $\omega$ in \eqref{DSMC: the effective diameter, calculation, final} determines the power law between viscosity coefficient $\mu$ and the temperature $T$ in the form $\mu \propto T^\omega$ \cite[Eq 3.66]{Bird1994MolecularVisc}. For Maxwell molecules, the viscosity is proportional to the temperature hence we use the $\omega=1$ in our DSMC simulation. 

We compute the upstream/downstream viscosity coefficient and heat conduction coefficient of our DSMC simulated gas using the Chapman-Enskog theory \cite[Eq 3.73]{Bird1994MolecularVisc}
\begin{equation} \label{DSMC viscosity coef}
\begin{split}
    	\mu &= \frac{5(\alpha+1)(\alpha+2)(\pi m k_B)^{1/2}(4k_B/m)^{\omega-1/2} T^\omega}{16\alpha\Gamma(\frac{9}{2}-\omega)\sigma_{T,ref}v_{r,ref}^{2\omega-1}}\\
	\kappa &= \frac{15 k_B}{4 m}\mu_0
\end{split}
\end{equation}
in which the reference total cross section $\sigma_{T,ref} =\pi d_{ref}^2 $ and the reference velocity $v_{r,ref}  = \sqrt{\frac{4k_B T_{ref}}{m\Gamma(5/2-\omega)^{\frac{1}{\omega-1/2}}}}$. 

The mean free path of gas from the collision rate per gas molecule are hence computed by \begin{equation} \label{Shock Wave Structure: mean free path}
	l= \sqrt{\frac{ \pi}{2 }}\frac{\mu}{\rho \sqrt{\frac{ k_B T}{ m}}}	
\end{equation}
in which $\rho$ is the mass density. Correspondingly, the mean free time of gas molecules is $t_m = l_m/\sqrt{\frac{8 k_B T}{\pi m}}$.

The flow profiles are sampled from simulation particles in each cell and averaged with previous steps every 4 time steps, while results from first 2000 time steps are discarded. The time step $\Delta t$ in our DSMC computation is choosen to be the $0.02$ times of the minimum mean free time between upstream and downstream mean free times, while the total simulation time is determined manually by checking the convergence of the simulation.

\paragraph{Declaration of generative AI and AI-assisted technologies in the writing process} During the preparation of this work the author(s) used ChatGPT for grammar checking. After using this tool/service, the author(s) reviewed and edited the content as needed and take(s) full responsibility for the content of the publication.

\bibliographystyle{elsarticle-num} 
\bibliography{refer}








\end{document}